\documentclass[twocolumn,amssymb,floatfix,aps,prb,showpacs,citeautoscript]{revtex4}
\usepackage{graphicx}
\usepackage{bm}
\usepackage{epsfig}
\usepackage{epstopdf}
\usepackage{xcolor}
\usepackage{hyperref}
\hypersetup{colorlinks=true,allcolors=blue}
\usepackage{natbib}
\usepackage{mathtools}
\usepackage{siunitx}
\usepackage{soul,xcolor}
\usepackage{amsmath,amssymb}
\graphicspath{{figures/}}

\setstcolor{red}
\DeclareUnicodeCharacter{2212}{-}
\DeclarePairedDelimiterXPP\BigOSI[2]%
  {\mathcal{O}}{(}{)}{}%
  {\SI{#1}{#2}}

\begin{document}
\title{ Quantum Fidelity of the Aubry-Andr\'e Model and the Exponential Orthogonality Catastrophe}

\author{J. Vahedi}
\email[]{j.vahedi@jacobs-university.de}
\affiliation{Department of Physics and  Earth Sciences, Jacobs University Bremen, Bremen  28759, Germany}  

\author{S. Kettemann}
\email[]{s.kettemann@jacobs-university.de}
\affiliation{Department of  Physics and Earth Sciences, Jacobs University
  Bremen, Bremen 28759, Germany}  
\affiliation{Division of Advanced Materials Science, Pohang University of Science and Technology (POSTECH), Pohang 790-784, South Korea}

\begin{abstract} 
 We consider the   orthogonality catastrophe
in the    (extended)
  Aubry-Andr\'e (AA)-Model, by calculating  
 the  overlap $F$
    between the ground state 
      of  the Fermi liquid in that quasi-crystalline model and the one 
       of the same system with an added potential impurity, as function of the size of 
        that impurity.
    Recently, the typical fidelity $F_{\rm typ}$   was found   in  quantum critical phases
     to decay 
exponentially with system size $L$  as $F \sim \exp(-c L^{z \eta})$\cite{Kettemann2016}
as found in an analytical derivation 
due to critical correlations. 
 For the critical AA model $\eta  = 1/2$
   is the power of multifractal intensity correlations, and $z$ the dynamical exponent due to the fractal structure of the density of states which is numerically found to be $z \gg 1$.
   Therefore, 
    we aim here to check this prediction by numerical finite size scaling. Surprisingly, however, 
     we find for a  weak single site impurity that the fidelity decays with a power law,  in the  critical phase. Even though it is found to be  smaller and decays  faster than in the metallic phase, it does not decay exponentially as predicted. We find an 
    exponential AOC however in the insulator phase
    for which we give a
    statistical explanation, a mechanism which is  
     profoundly 
    different from the AOC in metals, where it is the coupling to a continuum of states which yields there the power law suppression of the fidelity. 
       By reexamination of the analytical derivation we identify nonperturbative corrections due to the impurity potential and  multipoint correlations among wave functions as possible causes for the absence of the exponential AOC in the critical phase. 
    For an extended impurity, however, we find indications of  an 
     exponential AOC at  the quantum critical point of the AA model and at the mobility edge of the extended AA model and suggest an explanation for this finding. 
     Furthermore we consider a  
     parametric perturbation to the AA model, and find an exponential AOC numerically, in agreement with an analytical derivation 
      which we
     provide here. 
\end{abstract} 

\maketitle

\section{Introduction} 
\begin{figure}
\centering
\includegraphics[width=0.5\textwidth]{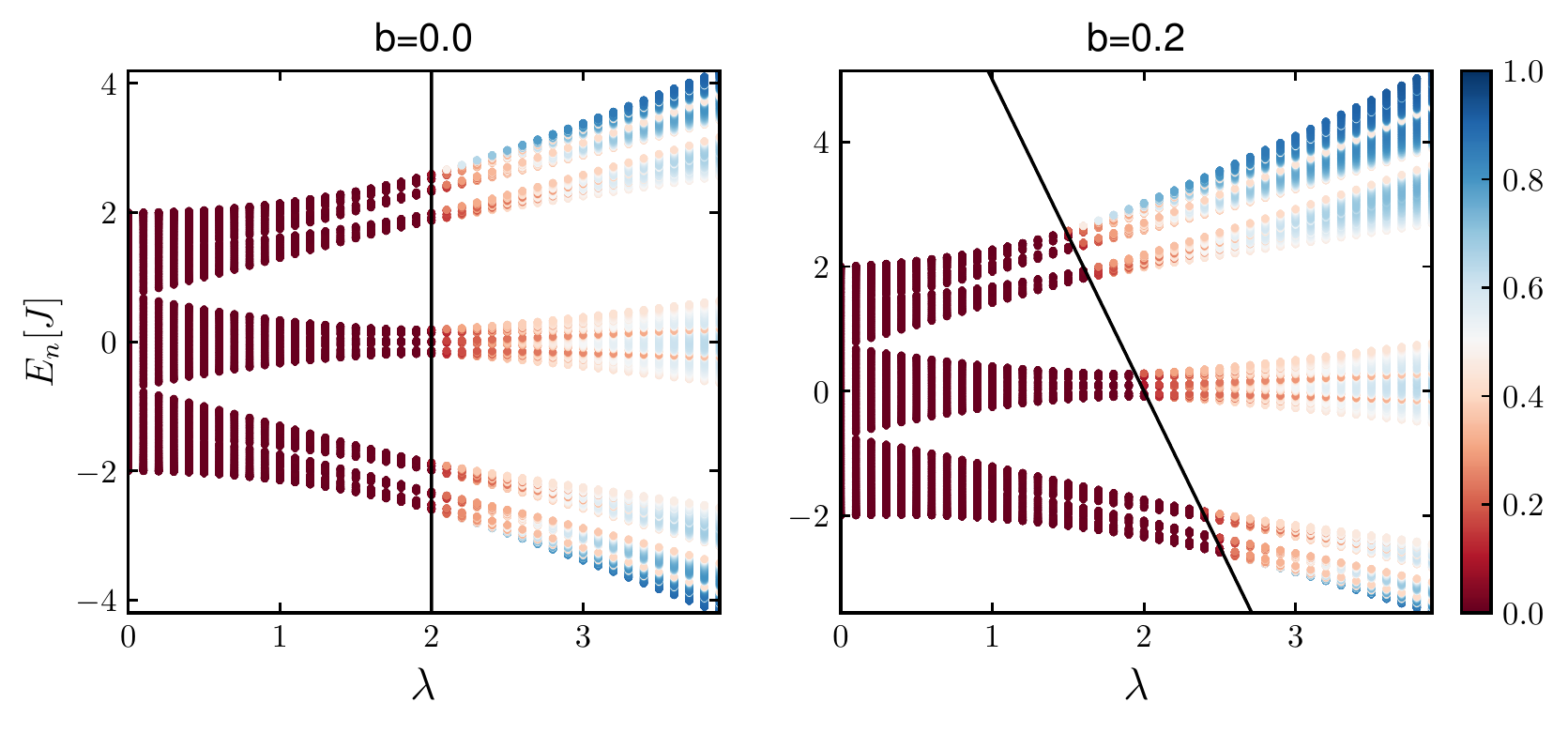}
 \caption{Inverse participation ratio ${\rm IPR =\sum_i|\psi_i|^4}$ as function  of  energy   and parameter $\lambda$ for the  AA-model (left) , and the  EAA-model with $b=0.2$ (right), where a mobility edge (black solid line) separates the extended phase ${\rm IPR}\rightarrow0$ from the localized 
 ${\rm IPR}\rightarrow a/\xi$, where $\xi$ is the localization length and $a$ the lattice spacing. System size is $L=610$.}
\label{ipr}
\end{figure}
The quantum fidelity $F$, the absolute value of the  scalar product between 
 the ground state of a quantum system  $|\psi>$ and the ground state after 
  a perturbation  $|\psi'>$, at fixed density  of fermions $n=N/L^d$,
 $F=|<\psi |\psi'> |$  is known to vanish with a power law of the system size $L$ in a metallic  phase, the celebrated Anderson orthogonality catastrophe\cite{ao}. 
 Anderson showed  in Ref. \onlinecite{ao}
 that  
   the fidelity  has a strict upper bound,
   \begin{equation} \label{inequality}
    F =
     |\langle  \psi \mid \psi' \rangle| < \exp \left(- I_A\right),
   \end{equation}
   where the Anderson integral $I_A$ is 
    for noninteracting electrons given in terms 
     of the single particle eigenstates of the original system 
     $|n\rangle$ and the new system $|n'\rangle$ by 
      \begin{equation} \label{IA}
I_A = \frac{1}{2} \sum_{n = 1}^{N} \sum_{ n' > N}
|\langle n | n' \rangle|^2.
            \end{equation} 
 If the added   impurity is short ranged and 
  of strength $V_0$, Anderson found 
   for a clean metal $I_A = (1/2) \rho_0^2 V_0^2 \ln N,$ where $\rho_0$ is the density of states at the  Fermi energy,
 diverging with  the number of Fermions $N$, so that $F$ decays with a power law with  $N = n L^d,$    the so called Anderson
  orthogonality catastrophe (AOC). 
 According to  Eq. (\ref{IA})
  this suppression 
   is a consequence of the fact 
  that the local perturbation 
  connects the Fermi liquid to a
  continuum  of excited states. This  has 
  therefore  important experimental consequences like the  singularities in X-Ray absorption and emission of metals\cite{nozieresdominicis}.   Furthermore,  the zero bias anomaly  in
   disordered metals\cite{altshuleraronov} and 
   anomalies in the tunneling density of states in quantum Hall systems\cite{tunnelingDOS}
   are related to the OC.
   The concept of fidelity can be generalized
    to any parametric perturbation of a system and be used to 
     characterize quantum phase transitions \cite{venuti}. A relationship between the orthogonality catastrophe and the adiabaticity breakdown in a driven many-body system has been shown in 
     Ref. \cite{Cheianov2017}. 
   The orthogonality catastrophe  
       can be studied in ensembles of ultracold atoms in a
      controlled way\cite{demler}.
    \begin{figure*}[t]
\centering
\includegraphics[width=1.0\textwidth]{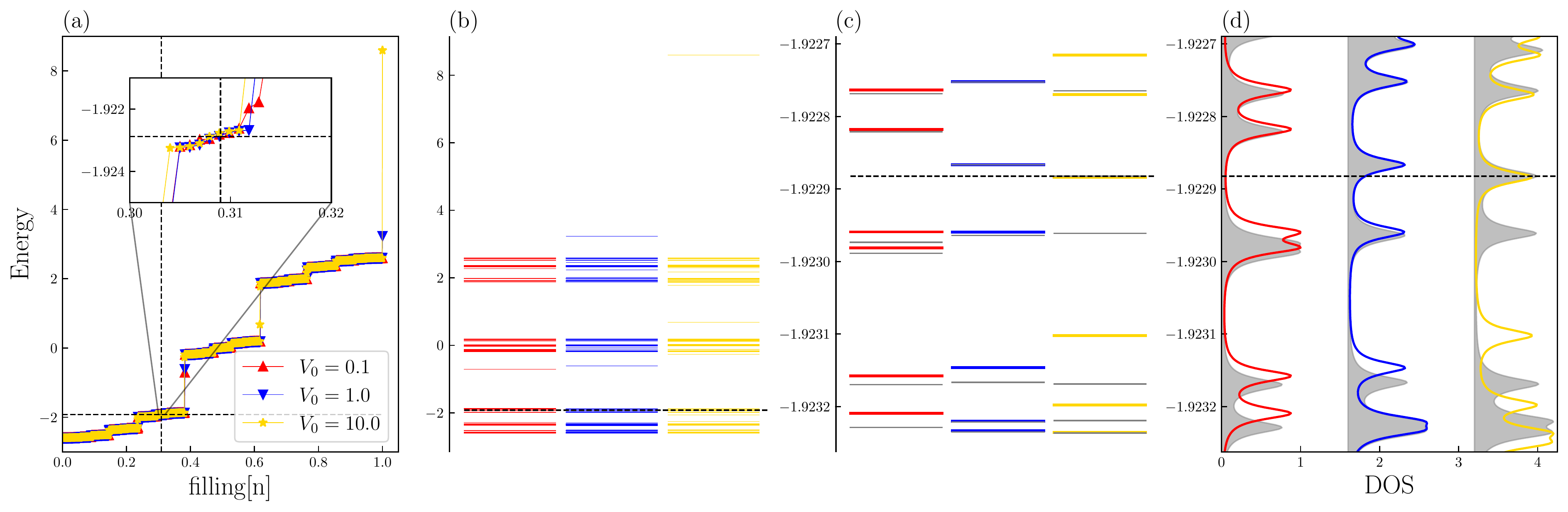}
 \caption{
(a)  Energy spectrum at the critical point $\lambda=2.0$ as  function of filling $n$ for a single impurity case. Results of three different impurity strength $V_0$ are displayed in  colored symbols. The horizontal dashed line is the Fermi energy. Inset: zoomed energies close to the Fermi energy without the impurity $E_F/J=-1.922882$, corresponding to filling $n = 0.309$. (b) and (c) show   the  complete  energy diagram, and a zoom close to the $E_F$,  with and without (grey) impurity. (d) density of state (DOS) as  function of energy close to  $E_F$. The results are randomly chosen from one of the realizations. For calculating DOS broadening $\eta=1e^{-5}$ is utilized. System size $L =1024$.}
\label{dos}
\end{figure*}

     Recently, it has been found that  the AOC with a local impurity  
     can be exponential  at  any quantum critical point, as  obtained in an 
      analytical derivation \cite{Kettemann2016}. There, the coupling to a continuum of excited states due to the impurity was found to be enhanced by  quantum critical power law correlations. 
      On the other hand, it has been argued 
      in Refs. \onlinecite{gefenlerner} and  \onlinecite{Kettemann2016} that in an Anderson insulator the fidelity with a local impurity remains  typically finite since the impurity can couple only to a discrete number of states. In Refs. \onlinecite{Khemani,Deng}, however, 
       an exponential AOC was found numerically in Anderson-localized Fermi systems, when the perturbation is turned on adiabatically slowly.
 
 In order to clarify the existence of an exponential AOC, here we  aim to study  
 the fidelity in the (extended) quasicrystalline Aubry-Andr\'e-Model\cite{AA1980,Hiramoto1992,Ganeshan2015}.
 The AA-model 
  has a quantum phase transition from a metal to a localised phase as function of a parameter $\lambda$ and  a quantum critical point $\lambda=2$, see Fig. \ref{ipr} (left).
 In a recent study \cite{Wu2019} the AOC at the critical point of this model has been  studied and found numerically to follow a power law. This is an another motivation for us, to reconsider the 
  fidelity in this model and to examine whether the exponential AOC predicted in Ref. \cite{Kettemann2016} at a quantum critical point exists in the critical AA-Model.
 Moreover, this model and its extensions can be realised in ultracold atoms, allowing the tuning of parameters and perturbations in a controlled way \cite{demler}.
  
The  (extended) Aubry-Andr\'e-Model has the  Hamiltonian\cite{AA1980},\cite{Ganeshan2015}
\begin{eqnarray}
 H_{EAA} =&& -J \sum_{i=1}^{L} (c_i^+ c_{i+1} + c_{i+1}^+ c_{i}) \nonumber\\
 &&+ \lambda \sum_{i=1}^L \frac{\cos (2 \pi Q i +
\phi)}{1-b\cos (2 \pi Q i +
\phi)}  c_i^+ c_{i},  
\label{eqEAA}
\end{eqnarray}
  where  $J$ is the hopping amplitude (we set $J=1$ as the unit of energy), $c_i^\dagger$ and $c_i$ are creation and annihilation
operators of a spinless fermion at site $i$ on 
 a chain of  $L$ sites,
 and $\lambda$ presents the amplitude of the
quasiperiodic potential. $Q$ is an irrational number usually chosen to be the golden ratio, $Q=2/(\sqrt{5}+1)$, and $\phi$ is a randomly chosen phase interval $[0,2\pi]$  that is the same for all sites. The open boundary conditions are considered throughout the results presented in the paper.
The parameter  $b$ can take values  $b\in[0,1)$. For $b=0$ we recover 
 the Aubry-Andr\'e-Model, which  has no mobility edge in the energy spectrum, 
 but when  the parameter $\lambda$  is changed all states undergo 
 a transition from  localized $\lambda>2$,
critical $\lambda =2$ to  extended for  $\lambda<2$ \cite{AA1980}, as seen in Fig. \ref{ipr} (left), where  the inverse participation ratio (${\rm IPR =\sum_i|\psi_i|^4}$) is
 plotted versus   energy   and parameter $\lambda$. At the critical point  $\lambda_c =2$,  all 
 eigenstates  are known to be  multifractal\cite{Hiramoto1992}.
 Moreover the model has  a fractal energy spectrum, where the level spacing $\Delta$ scales
  with system size $L$ as  $z$, $\Delta \sim L^{-z},$ where the dynamical exponent $z$ can be different from the dimension of the model $d=1.$
  
  For $b\neq 0$ the EAA-model shows a mobility edge given by $E_{\rm mb}=(2J-\lambda)/b$\cite{Ganeshan2015}, as seen in Fig. 
  \ref{ipr} (right), where  the inverse participation ration (IPR) 
   is plotted as function of energy and parameter $\lambda$. 
   The  mobility edge (black solid line) separates the extended phase ${\rm IPR}\rightarrow0$ from the localized 
 ${\rm IPR}\rightarrow a/\xi,$ where $a$ is the lattice spacing and 
 $\xi$ the localization length. 
  
 The paper is organised as follows. In section II
we review the 
definition of the ground state fidelity $F$ in the presence of an impurity  and its upper bound 
provided by the exponential of the 
Anderson integral. In section III we define the Anderson integral with  an extended impurity. 
In section IV we review the spectrum of the AA model, and study how it is modified by an  impurity. 
In section V we present all results for the fidelity of a single site  impurity in the AA model. 
We begin with presenting the numerical results in section V. A. The analytical results in the approximation used in Ref.~\cite{Kettemann2016} are reviewed and applied for the AA model with a single site impurity in section V. B yielding an exponential 
 AOC in the critical phase. As this is in disagreement with the numerical results presented in section V. A,
 we consider corrections to the Anderson integral in section V. C beyond 
 the approximation used in 
 section V. B. We thereby succeed to  identify a mechanism which yields a potential AOC in the critical phase, in agreement with the  numerical results.  In section V. D we show that in the insulator regime there is a statistical  mechanism which yields an exponential AOC in agreement with the numerical results in the insulator phase of the AA model. In section VI we present results for the fidelity with an extended impurity, and provide evidence for  an exponential AOC in the critical regime, when the impurity extends over more than one sites. By analysing the Anderson integral for an extended impurity we suggest a mechanism which  explains this discovery of an exponential AOC in the critical phase. In section VII we present numerical results for the  ground state fidelity with a parametric perturbation giving evidence for an exponential AOC. Analysing the Anderson integral for that perturbation we give a derivation which is in agreement with these numerical results. In section VIII we present numerical results for the  ground state fidelity in the extended AA model. 
 We conclude with section IX.  In Appendix A we review the  derivation of  an upper bound for  the ground state fidelity. In Appendix B we give details for the derivation of the AOC with a single site impurity. In Appendix C we present numerical benchmark results for the fidelity of the 1D tight binding model with an impurity.

\section{Ground State Fidelity}

To derive the ground state fidelity 
we first
diagonalise 
 the Hamiltonian  as  given by Eq. (\ref{eqEAA}). As the model is  non interacting, it can be diagonalised  with the basis change as.
 $H_{AA}=\sum_{n}
 \epsilon_{n} 
 d^{\dagger}_{n}
 d_{n}$, with the one-electron energy Eigenvalues $\epsilon_{n}$, and 
  the creation and annihilation operators in the  single particle Eigenstates $|n \rangle$ given by $d_{n}=\sum_i \psi_{n i} c_i$, where $\psi_{n i}$ are complex coefficients. Then, the ground state can be constructed as 
  $|\psi \rangle=\prod_{n =1}^N d^\dagger_{n} |0\rangle$, 
  with  fixed   number of particles $N$ and 
 fixed   particle filling $n =N/L$. 
When adding a perturbation the filling $n$ remains fixed, while  the   Fermi energy can change.

Next,  we introduce an  impurity, which  extends over a finite subset $S_M$ 
of $M$ neighboured  lattice sites with  $S_M = i,i+1,..i+M-1$ 
with 
\begin{equation} \label{imp}
H_{imp} =  \frac{1}{M} V_0 \sum_{i \in S_M} 
c_{i}^{+}\,c_{i}.
\end{equation}
In the numerical implementation we  choose the
center of the impurity to be located at  the 
lattice center $L/2$. 
The non interacting Hamiltonian  perturbed by the impurity  $H'=H_{AA}+H_{imp}$ 
  has the new 
 Eigenstates $|n' \rangle,$ yielding
$H'=\sum_{n'}\epsilon'_{n'}d'^\dagger_{n'}d'_{n'}$, where $d'_{n'}=\sum_i \psi_{n' i}c'_i$,
with complex coefficients $\psi_{n' i}$. 
Thereby, the new ground state  is given by
$|\psi'\rangle=\prod_{n' =1}^N d^\dagger_{n'}|0\rangle$. 
 Thus,  the fidelity is  given by
$F= |\langle\psi'|\psi\rangle|
=|{\rm det}(A)|$, where $A$ is the $N \times N$  matrix where the matrix elements are the scalar products of the Eigenstates before and after the perturbation,  $A_{n n'} = \langle  n | n' \rangle $, see Appendix A for more details.

\section{Anderson Integral}
 
 Before presenting the numerical results, let us first review the rigourous 
 upper limit of the fidelity, as given by the right hand side of Eq. 
  (\ref{inequality}), whose derivation is given in Appendix A. 
 The Anderson integral Eq.   (\ref{IA})
can  be rewritten for the  impurity perturbation Eq. (\ref{imp}) 
 without approximation
as 
   \begin{eqnarray} \label{ia}
I_A  &=& 
       \frac{1}{2} \sum_{n=1}^N \sum_{n'>N}
   \frac{1}{(E_{n'}-E_n )^2}  |  \langle n | H_{imp} | n' \rangle|^2  \nonumber \\ &=& 
       \frac{ V_0^2}{2 M^2} \sum_{n=1}^N \sum_{n'>N}
   \frac{ | \sum_{i \in S_M } \psi_{n i}^* \psi_{n' i} |^2}{(E_{n'}-E_n )^2} , 
  \end{eqnarray}
  where $ \psi_{n i} = \langle n | i \rangle$,  
  $ \psi_{n' i} =  \langle n' | i \rangle$, is the local amplitude 
   with and without the additional impurity at site $i.$ 
   Eq. (\ref{ia}) can be rewritten by replacing the summation over energy Eigenvalues $E_{n'}, E_n$ 
    by an integral over energy with density of states $\rho(E)$ without the 
    impurity, and $\rho'(E')$ with the impurity.
    Thus, we get 
       \begin{eqnarray} \label{ia2}
I_A  &=& 
      \frac{ V_0^2}{2 M^2} 
        \int_{E \le \epsilon_{HOMO}} d E \int_{E' \ge \epsilon_{LUMO}'} d E'
         \frac{\rho (E) \rho'(E')}{(E_{n'}-E_n )^2}
        \nonumber 
    \\    &\times&
    \sum_{i,j \in S_M } \psi_{E i}^* \psi_{E' i}  \psi_{E j} \psi_{E' j}^*, 
  \end{eqnarray}
  which depends explicitly both on the density of states (DOS) with and without impurity,$ \rho'(E'),\rho(E)$ and on the wave function amplitudes with and without the impurity
  $\psi_{E' i},  \psi_{E i}$.
  We note that, since the number of fermions $N$ is kept fixed, the Fermi energy of the pure system
  $\epsilon_F  $  can be different from the one of
   the system with the perturbation $\epsilon_F'$,
    since all energy levels $E_{n'}$  may change
 with the perturbation. 
    We therefore find it convenient to  define the 
    highest occupied energy level without the perturbation as $\epsilon_{\rm HOMO}$
    and the lowest unoccupied energy 
     level with the perturbation as
      $\epsilon_{\rm LUMO}'.$
    We note that Eq. (\ref{ia2}) is still an exact representation of the Anderson integral, rewritten in terms of the density of states.. 
   As the density of states  of the AA model is known to show fractal behavior at the critical point $\lambda_c=2,$ let us first  consider the effect of the impurity on the energy spectrum. 
    
   \section{Energy Spectrum} \label{energyspectrum}
     In Fig. \ref{dos} a)  we show   the 
     energy level spectrum of the AA model, Eq. (\ref{eqEAA}) for $b=0$
   and the critical parameter $\lambda_c=2.0$
   as function of filling factor $n$. 
     The  dashed line indicates the 
   filling of $n = 0.309$,  
   corresponding without an impurity  to the Fermi energy $E_F/J=-1.923$. In the  inset  the zoomed energy interval  close to  that filling $n = 0.309$ is seen to correspond to  a region of large density of states. For a single site  impurity the energy level spectrum is plotted  for three different impurity strength $V_0$ as displayed by the   colored symbols, respectively.  Fig. \ref{dos} (b) and (c) show a full and zoomed energy level diagram, with and without impurity. The case without impurity is drawn in a grey color. While the energy bands are not shifted, the formation of bound states outside of the energy bands is seen even for the weakest impurity strength. Fig. \ref{dos}  (d) shows the density of state (DOS) as a function of energy close to Fermi energy. The results presented here are randomly chosen from one of the realizations. For  the calculation of  the  DOS a broadening $\eta = 1. \times e^{−5}$ has been used. 
   
   Fig. \ref{dos_ave} shows the average density of states as function of energy $E$, as averaged over the random phases
 $\phi$ in the Hamiltonian Eq. (\ref{eqEAA}) for $b=0$ and $\lambda=2$ of 200 realizations.
  This supports  the observation that the 
  energy bands are not shifted by more than a level spacing, and that the formation of bound states outside of the energy bands is seen even for the weakest impurity strength.

\begin{figure}
\centering
    \includegraphics[width=0.5\textwidth]{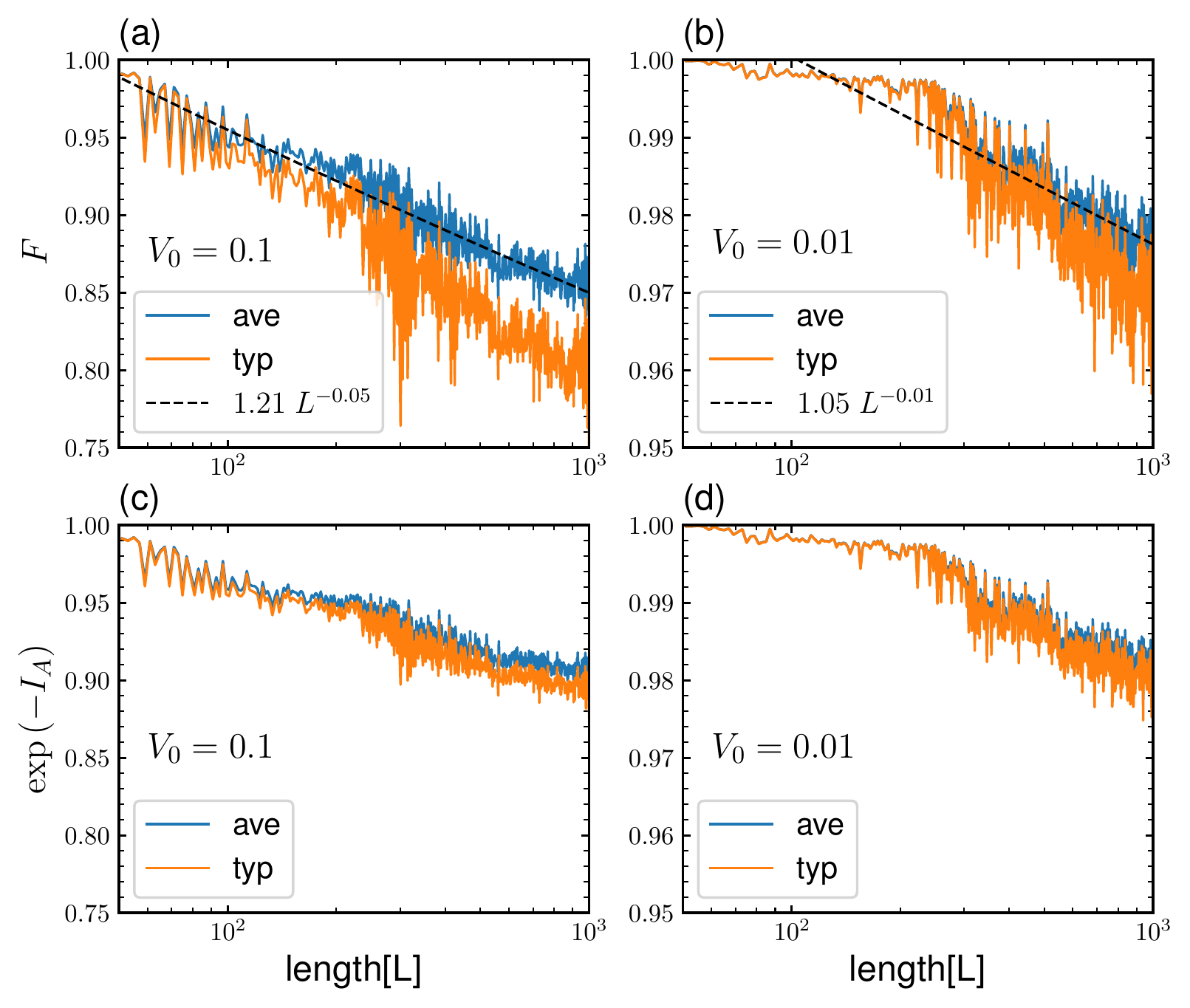}
 \caption{ (a) Average and typical  fidelity  as function of  length $L$  with a single impurity of strength  $V_0=0.1$   and (b) for  $V_0=0.01$.  The dashed line is a fitted power law as given in the legend. (c) Average and typical value of  $\exp (-I_A)$, where $I_A$ is the  Anderson integral  Eq.~(\ref{ia}) for $V_0=0.1$   and (d) for  $V_0=0.01$. Parameter $\lambda_c=2.0,$ data   averaged over $1000$ sample realizations. }
\label{fig:singlesite}
\end{figure}

  \section{Ground State Fidelity of the AA-Model with a Single Site Impurity}
  
   \subsection{Numerical Results}

  Let us first consider  the fidelity in the 1D Aubry-Andr\'e model
  with Hamiltonian Eq. (\ref{eqEAA}) for $b=0$,
  for a single site impurity, $M=1$, Eq.(\ref{imp}), numerically. Calculating the fidelity, using its  definition $F= |\langle\psi'|\psi\rangle|$, we
   plot in Fig. \ref{fig:singlesite}  (a), (b)
  the average  fidelity  $F_{ave}=\left<F\right>$ and 
   the typical fidelity as function of length $L$ for different impurity strengths $V_0 = 0.1,0.01.$
  Here,  we defined  
  $F_{typ}=\exp{\left<\log{F}\right>}$, where $\left<\dots\right>$  denotes the  average over $1000$ realizations 
  of a uniform random phase in $[0,2\pi)$.
  We find that both the average and the typical fidelity decay with a power law, and not exponentially. 
 The typical fidelity is smaller 
 than the average one 
 for all system sizes  $L.$ This difference becomes more pronounced
    with stronger impurity strength $V_0,$ while the fidelity 
    becomes smaller with increasing $V_0$
    overall. 
    For comparison, we also calculated the Anderson integral $I_A$, and plot in Fig. \ref{fig:singlesite} (c), (d)
     the average of its  exponential  $\langle \exp (- I_A) \rangle$, which should give 
     according to Eq. (\ref{inequality})
     the upper bound of 
      the average fidelity, 
     and  the exponential of the average $I_A$, $\exp(-\langle I_A \rangle)$,  which corresponds to the upper bound for 
      the typical  fidelity. Indeed, we confirm the inequality Eq. (\ref{inequality})
      for both impurity strengths  $V_0 = 0.1,0.01$. But  we observe that 
       the typical  fidelity  $F_{\rm typ}$ is substantially smaller than its  upper bound
       $\exp(- \langle I_A \rangle)$.

     \subsection{Anderson Integral: Analytical Results}   
        \label{sec_aianalytical}

These numerical results are in contradiction with 
 the  prediction of an exponential orthogonality catastrophe, as found by an 
      analytical derivation  in Ref. \cite{Kettemann2016}
      at a quantum critical point, where the coupling to a continuum of excited states due to the impurity was found to be enhanced by  quantum critical power law correlations. 
       Let us therefore reconsider the derivation of the 
 Anderson integral for critical states.  
  
   In  fact, in the critical regime 
    all wavefunctions are multifracral\cite{multifractal}
    and the 
    correlation function of 
intensities associated to two energy levels distant in energy  by
$\omega_{nm} =E_n- E_m$ is enhanced, as  given by
\cite{powerlaw,cuevas,ioffe} 
\begin{eqnarray}
\label{cc}
 && C(
\omega_{nm} = E_n-E_m) = L^d \int d^dr\, \left\langle |\psi_n({\bf r})|^2
 |\psi_m({\bf r})|^2 \right\rangle \nonumber \\ &&= \left\{
 \begin{array}{ll}(\frac{E_c}{{\rm Max} (|\omega_{nm}|, \Delta) })^{\eta/d}, 
 & 0 < |\omega_{nm}| < E_c, \\ (E_c/|\omega_{nm}|)^{2}, &
|\omega_{nm}| > E_c,
\end{array}
\right.,
\end{eqnarray}
where $\Delta$ is 
the average level spacing  at the Fermi energy.
The power is given by  $\eta = 2 (\alpha_{0}-d)$, with multifractality parameter $\alpha_0$  
 and the dimension $d$ . 
 For the critical AA model, the power is known to be
          $\eta = 1/2$ \cite{Duthie2021}. Since
 all its states  are critical, 
 the correlation energy $E_c$ is  of order
  of   the band width $D.$ 
For
$|\omega_{nm}| < E_c$ correlations are  thus indeed enhanced in comparison to the
plane-wave limit $C_{nm}=1$. Note, that for $|\omega_{nm}| > E_c$
 it decays below $1$.

          {\it Mean Value of  the Anderson Integral.---}
         If we  assume that   the perturbed eigenstates $\langle n' | $ in Eq.  (\ref{ia}) 
  can be replaced 
        by an eigenstate without the impurity,  $ \langle n | $, we can 
          insert  the correlation function Eq. (\ref{cc}) into Eq. (\ref{ia})
          to calculate 
         the mean value of  $I_A$, and find  
             for a single site impurity, $M=1$,
                  \begin{equation} \label{iafm}
       \langle  I_A \rangle = \frac{ V_0^2}{2 }
       \hspace{-.5cm}\iint\limits_{\epsilon < \epsilon_{\rm Homo},
        \epsilon' > \epsilon_{\rm Lumo}  }\hspace{-.5cm}d \epsilon d \epsilon'
         \rho(\epsilon) \rho(\epsilon') \frac{C_{\epsilon,\epsilon'}  }{(\epsilon-\epsilon')^2}.
         \end{equation}
          This  gives an estimate for the upper bound of the  typical  average of $F,$
         $\exp (\langle \ln F \rangle)\le \exp (-   \langle  I_A \rangle  )$.
         Assuming furthermore   that the  density of states is only slowly varying
         $\rho(E) \approx \rho (E_F) = \rho_0$,
         and denoting 
         the level spacing at the Fermi
         energy $\Delta =  \epsilon_{\rm Lumo} -  \epsilon_{\rm Homo}, $
         we get at 
         the AMIT
           with Eq. (\ref{cc}) 
           \begin{equation} \label{iafmEM}
       \langle  I_A \rangle|_{ E_F = E_M} = \frac{(\rho_0 V_0)^2}{2 \gamma (1+\gamma) }
      \left(  \frac{E_c}{ \Delta} \right)^{\gamma},
         \end{equation}
        depending on    $ E_c/\Delta$
          with  power  $\gamma = \eta/d$.  For the critical phase of the 1-dimensional  AA model, $d=1$,
          $\gamma = 1/2$\cite{Duthie2021}.

\begin{figure}
\centering
    \includegraphics[width=0.5\textwidth]{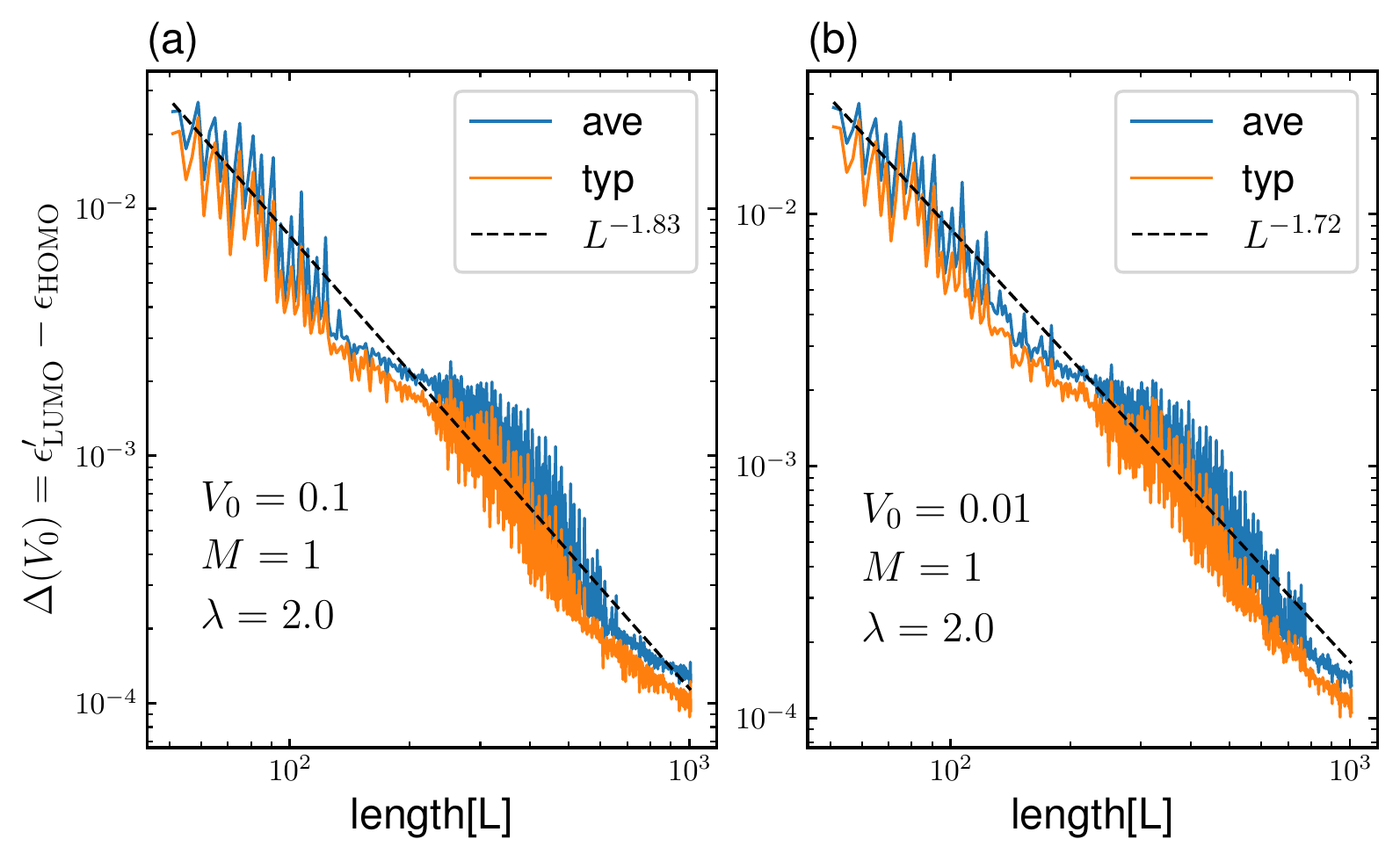}
 \caption{ Average and typical gap, the  difference between HOMO and LUMO energies  Eq. (\ref{gapv0}) of the model before and after quench were considered) for two different single impurity strengths (a) $V_0=0.1$  and (b) $V_0=0.01$. In all results, we fix parameter $\lambda=2.0$ and data averaged over $1000$ sample realizations.}
\label{gapsingle}
\end{figure}

         Since all states are critical at $\lambda =2$, we 
         set the correlation energy
          to  the band width $E_c=D$. In
           a metal the average level spacing  is
           $\Delta = 1(\rho_0 L) $.
           Note, however  that for the spectral spectrum of the AA model, 
          the 
          level spacing at the Fermi energy scales with $L$ rather as
          $\Delta (L)  \sim L^{-z},$ with $ z >1$\cite{Wu2019}.
          Thereby, we get 
          \begin{equation} \label{iafmEMAA}
       \langle  I_A \rangle|_{ E_F = 0} = \frac{ \rho_0^2 V_0^2}{2 \gamma (1+\gamma) }
      \left( D \rho_0 L^z \right)^{\gamma}.
         \end{equation}
          Thus, we get with $\gamma = \eta/d = 1/2,$ $\rho_0 =1/D,$ that the Anderson integral diverges as a power low with system size $L$
             \begin{equation} \label{iafmEMAA2}
       \langle  I_A \rangle|_{ E_F = 0} = \frac{ 2 V_0^2}{ 3 D^2  }
      L^{z/2},
         \end{equation}
    and thus the typical  fidelity decays exponentially with the system size, the exponential orthogonality catastrophe,
     in agreement with Ref. \cite{Kettemann2016}. 
    
      In the metallic regime 
      $\lambda <2$ one rather  gets 
      \begin{equation} \label{iafmEMAAM}
       \langle  I_A^M \rangle|_{ E_F = 0} = \frac{ \rho_0^2 V_0^2}{2 }
      \ln ( D \rho_0 L ).
         \end{equation}

      Thus, in order to be able to distinguish
      the exponential 
      decay of the typical fidelity due to critical correlations, Eq. (\ref{iafmEMAA2})
       in comparison to the noncritical result
       Eq. (\ref{iafmEMAAM}) the system size $L$
        should be so large that 
       $4/3 L^{z/2} > \ln L$   
      which is indeed  valid for all  $L >  1 $ for $z>1$. Thus, according to this result, 
       the numerical calculations should see the exponential decay, if the analytical 
       derivation  is valid. Therefore, let us reconsider the approximations yielding to the result Eq. (\ref{iafmEMAA}), in order to find out the reason for this discrepancy with the numerical results.

   \subsection{Anderson Integral: beyond perturbation theory.}
   
   1. As the energy levels are modified by the perturbation, the gap between the 
    lowest occupied state with the perturbation and the highest occupied level without perturbation depends itself on the disorder potential $V_0$, 
    \begin{equation} \label{gapv0}
    \Delta (V_0) =  \epsilon_{\rm LUMO}'-\epsilon_{\rm HOMO}.
    \end{equation}
    Since it provides the infrared cutoff to the integrals in the Anderson integral, 
    we thereby find  in the critical phase
   \begin{equation} \label{iafmEM}
       \langle  I_A \rangle|_{ E_F = E_M} = \frac{(\rho_0 V_0)^2}{2 \gamma (1+\gamma) }
      \left(  \frac{E_c}{ \Delta (V_0)} \right)^{\gamma},
             \end{equation}
     
     In Fig. \ref{gapsingle} the  average and typical gap
     $\Delta(V_0)$,
     Eq. (\ref{gapv0}) is shown
      for two different single impurity strengths, namely (a) $V_0=0.1$  and (b) $V_0=0.01$ for critical   parameter $\lambda=2.0$ and data averaged over $1000$ sample realizations.
We see that  the magnitude of the gap is not changed by the impurity, so that this weak  dependence of $\Delta(V_0)$
 on $V_0$, does not change the result for the Anderson integral Eq. (\ref{iafmEMAA2}).
 
 Also, as  was observed for the gap of the unperturbed system in Ref.  \cite{Wu2019}, 
 the decay with system size $L$ of the gap  $\Delta(V_0)$
 is strongly fluctuating with $L$  and does not follow
  a clear scaling law, even when averaging over 1000 realizations.  However, it clearly  decays 
  with a power
    $z>1$, faster than the average level spacing of a metal.

 2.   The density of states is affected by the presence of the impurity, 
 as seen in Fig. \ref{dos} for a particular realisation of the phase $\phi$. 
 As discussed in section \ref{energyspectrum},   the energy bands are hardly shifted, 
 but the formation of bound states outside of the energy bands is  found  even for the 
   weakest impurity strength. Close to the  Fermi energy
  Fig. \ref{dos}  (d) shows  that the density of state (DOS)  is only weakly 
 shifted  by the impurity, by the order of the level spacing $\Delta$.
 As we choose an energy region of large density of states, which is not shifted
 on average by the impurity, as seen  in
 Fig. \ref{dos_ave}, we conclude that the small change of DOS 
  $\rho'(E)$ by the impurity does not result into a change of the divergence with system size $L$
of the   average  Anderson integral in the critical regime, and thus cannot be responsible for the discrepancy with the numerical results.

3. We disregarded in the derivation in section   \ref{sec_aianalytical} the 
change of wave function intensity at the location of the  impurity by the addition of the impurity.
For a single site potential impurity at site ${\bf x }$
       with amplitude $V_0$,
        the perturbed intensity 
        $| \psi_{n'}({\bf x}) |^2$
        can be written exactly as\cite{Economou1983}
          \begin{eqnarray} \label{ec}
 |\psi_{n'} ({\bf x})|^2 = \lim_{ E \rightarrow E_{n'}}(E-E_{n'}) 
           \frac{V_0 (G_E^0({\bf x},{\bf x}))^2 }{1- V_0 G_E^0({\bf x},{\bf x})}, 
      \end{eqnarray}
      where, 
      \begin{equation}
      G_E^0({\bf x},{\bf x}) = 
      \sum_l|\psi_{l} ({\bf x})|^2   
      \frac{1}{E-E_l+i \delta}.
      \end{equation}
         Performing the limit 
         in Eq. (\ref{ec})
        with de l'Hospital, one finds
 
     \begin{eqnarray}
 |\psi_{n'} ({\bf x})|^2 =  |\psi_{n} ({\bf x})|^2
           \frac{( 1 + \frac{ E_{n'}-E_n }{ |\psi_{n} ({\bf x})|^2} \sum_{l \neq n}
       \frac{ | \psi_l({\bf x}) |^2 }{ E_{n'}-E_l  } )^2 }{1+\frac{ (E_{n'}-E_n)^2 }{ |\psi_{n} ({\bf x})|^2}\sum_{m \neq n}
       \frac{ | \psi_m({\bf x}) |^2 }{(E_{n'}-E_m )^2}}, 
      \end{eqnarray}  
      where $E_n$ is the energy level closest in energy to 
       the perturbed energy  $E_{n'}$.  It depends on the disorder potential only implicitly 
      through the Eigen energy of 
       the perturbed state $E_{n'}.$
      Since  $E_{n'}-E_n$  has a polynomial dependence on the disorder potential $V_0$, we can approximate it by  the leading term,  linear in $V_0$, $E_{n'}-E_n \approx V_0 |\psi_{n} ({\bf x})|^2$, yielding 
        \begin{eqnarray} \label{correction}
 |\psi_{n'} ({\bf x})|^2 \approx   |\psi_{n} ({\bf x})|^2
           \frac{( 1 + V_0 \sum_{l \neq n}
       \frac{ | \psi_l({\bf x}) |^2 }{ E_{n'}-E_l  } )^2 }{1+V_0^2 |\psi_{n} ({\bf x})|^2\sum_{m \neq n}
       \frac{ | \psi_m({\bf x}) |^2 }{(E_{n'}-E_m )^2}}. 
      \end{eqnarray}  
      Inserting this approximation into the Anderson integral,  we can check whether these corrections in $V_0$ change the divergence of the Anderson integral.
      In the metal phase, $|\psi_{n} ({\bf x})|^2 \sim 1/L$ and due to the asymmetry of the summations in the numerator of 
      Eq. (\ref{correction}), we find only weak corrections, which do not change the $\ln L-$dependence of the Anderson integral in the metallic regime. 
        In the critical regime, however, all wave functions are multifractal, so that the local intensity $ |\psi_{l} ({\bf x})|^2$ is widely distributed and may vary strongly  with energy $E_l$. Then, the corrections due to the summations  in Eq. (\ref{correction}) both in the numerator and nominator may yield finite results, especially when the intensity of the state  at the Fermi energy at the location of the impurity $|\psi_{n} ({\bf x})|^2$
  happens to be  smaller than in other states. Inserting        Eq.(\ref{correction})  into the Anderson integral, 
   we see that multi point correlations of the intensity arise even for the average Anderson integral. 
   
     \begin{figure}
\centering
\includegraphics[width=0.45\textwidth]{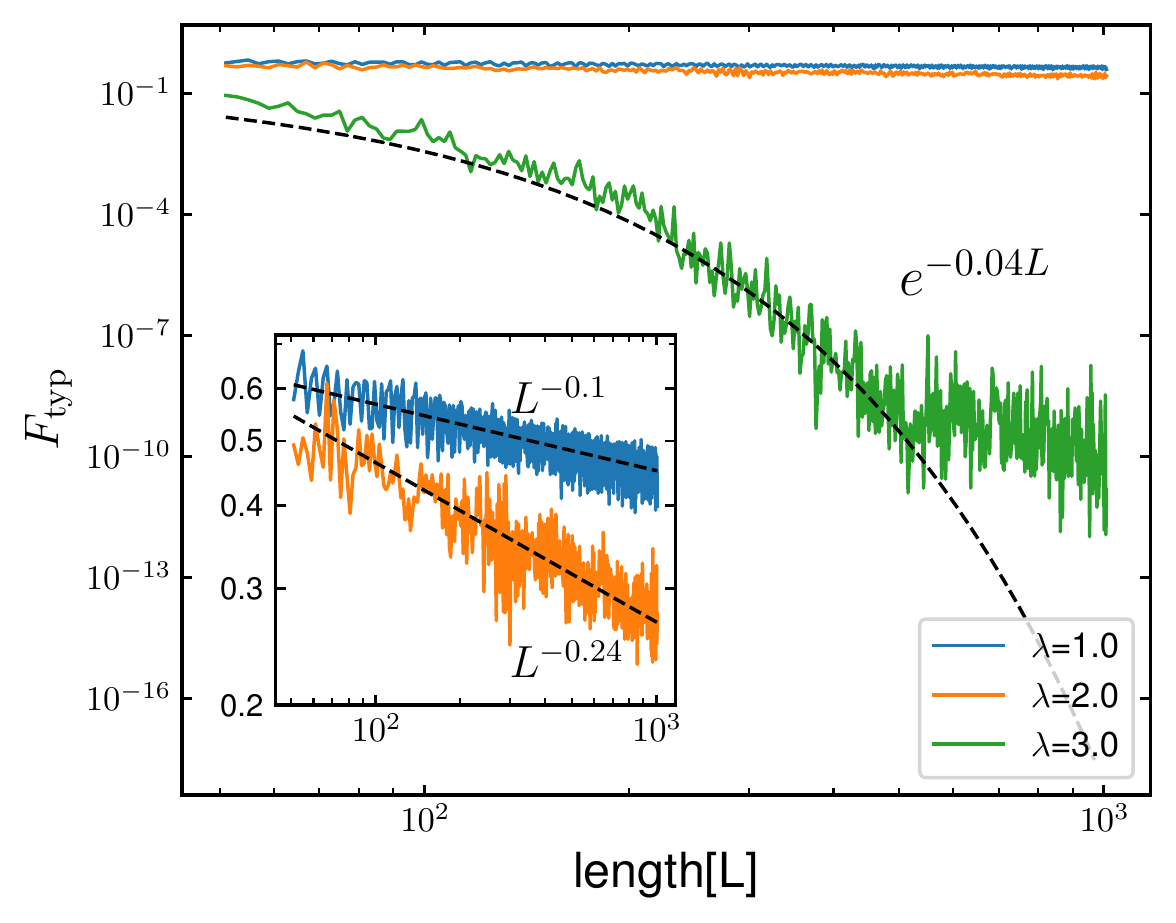}2 \caption{Typical fidelity $F_{typ}\approx\exp{\left<\log F\right>}$ of a single impurity with strength $V_0/J=20$ for different $\lambda$ parameter in $\log-\log$ scale. Filling fixed at $n=0.309$. Inset magnifies results of main panel for $\lambda=1.0, 2.0$.  Results are averaged over $1000$ samples.  Blake-dashed lines are fitted curves.}
\label{f_typVd20_m1}
\end{figure}

 \begin{figure*}
\includegraphics[width=1.0\textwidth]{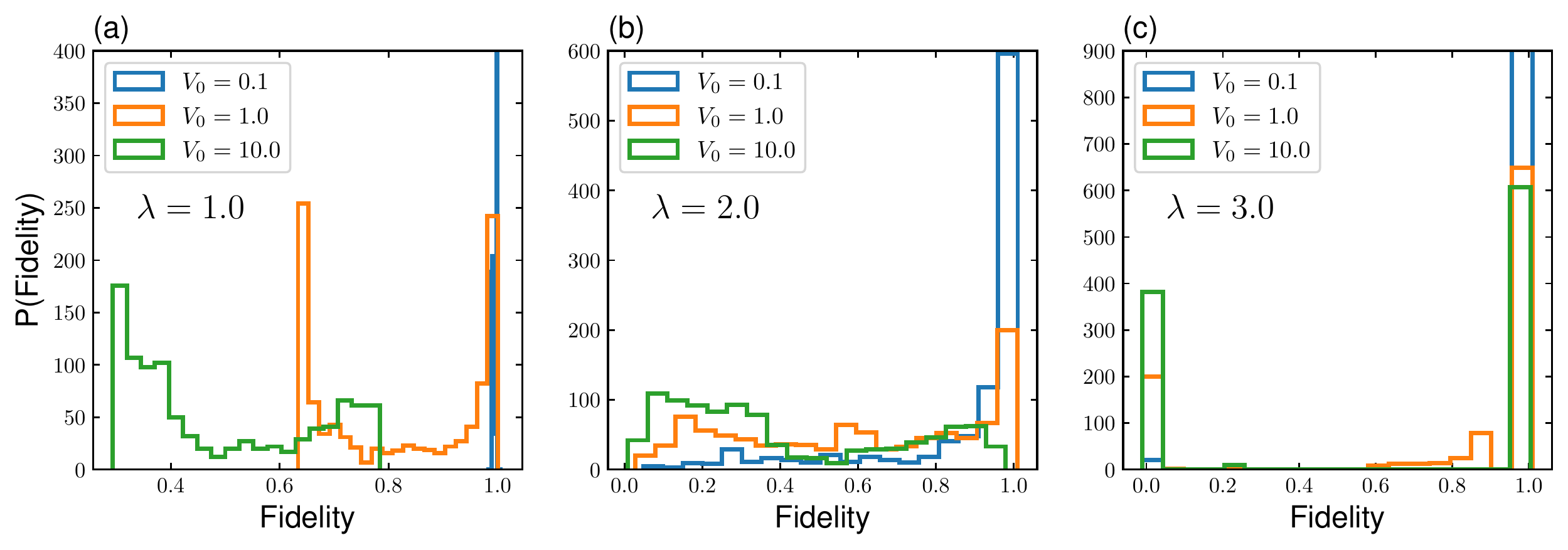}
 \caption{ Distribution of the fidelity for a  single impurity $M=1$ with different strengths. Panels are showing results for three (a) metallic, (b) critical, and (c) localized phases. System size is $L=1009$ and data are collected from $2000$ realizations. }
\label{distribution}
\end{figure*}
   
    Thus, the average  Anderson integral can in general not  be reduced to an integral over the  pair correlation function Eq. (\ref{cc}). The presence of multipoint correlation may therefore weaken the infrared divergence compared to Eq. (\ref{iafmEMAA2}). The numerical results shown in Fig.   \ref{fig:singlesite}   in fact provide strong evidence, that the Anderson integral depends on system size only logarithmically, resulting in a fidelity at the critical point which  decays with a power law with system size, albeit decaying faster than in the metallic regime. 
     Numerical Results for the fidelity in the presence of an impurity  in other quantum critical systems, in particular in  random banded  matrices \cite{Moure2017} and at the 3D Anderson metal-insulator transition \cite{Slevin2017} 
      did not find evidence for an  exponential AOC neither, 
      but rather found evidence for  a potential AOC. 
        As outlined above, the  explanation may be that  the corrections to the local intensity 
         at a single site impurity  Eq. (\ref{correction}) result in multipoint correlations, which  weaken the infrared singularity of the Anderson integral in these quantum critical systems, thereby explaining the  numerically observed power law Anderson orthogonality catastrophe.

 \subsection{Fidelity in the Insulator phase: 
 Statistical Exponential Orthogonality Catastrophe}
 \label{seoc}
 
 In Fig. \ref{f_typVd20_m1}  the typical fidelity $F_{typ} = \exp{\left<\log F\right>}$ of a strong single impurity with strength $V_0/J=20$  is shown for  the  metallic  $\lambda=1.0$,
  the critical 
 $\lambda_c=2.0$, and the insulator  $\lambda=3.0$ regime.
  The filling is kept fixed at $n=0.309$. All   results are obtained by  averaging over $1000$ samples. 
  We see that both in the metallic and the critical regime, the typical fidelity decays with system size like a power law, albeit
  the decay is faster in the critical regime. This is  seen better in the 
  inset where the dashed lines are  fitted curves as indicated. 

In the insulator regime $\lambda=3.0$, however, the fidelity is 
by orders of magnitude smaller than  that in the metallic and critical regimes. Moreover, it 
 decays  with system size $L$ exponentially, as seen by the fitted dashed line, until it decays more slowly at system sizes exceeding $L=500.$
In Refs. \onlinecite{Khemani,Deng},
       an exponential orthogonality catastrophe was found numerically in Anderson-localized Fermi systems, when the perturbation is turned on adiabatically slowly. In Ref.  \onlinecite{Khemani} that has been explained in terms of a {\it statistical} orthogonality catastrophe.
       
       Indeed, in the strongly localised regime, when each Eigenfunction is localised on one site only, one obtains a Bernoulli distribution of the fidelity of fixed particle number $N$, which is either $0$ or $1$ with probability $u,$ $1-u$, respectively. 
        The reason is, that in this strongly localised regime a local impurity cannot mix Eigenstates, but only shift the  energy of that state, which is located at the site of the impurity. Thereby, 
         for fixed number of particles the impurity may shift an occupied level 
          to higher energies, leaving it unoccupied, while a state at another site becomes occupied, which is orthogonal to the state at the site of the impurity, or vice versa. Thus, by definition of the fidelity at fixed $N,$ the fidelity is then exactly zero. 
           If, on the other hand, the impurity shifts the energy such that the level remains occupied when it was occupied before, or leaving it unoccupied, when it was unoccupied without the impurity,  the fidelity remains exactly one. 
          Thus, one has a statistical distribution which has only two possibly values, $S=0$ with probability $u$ or $S=1$ with probability $1-u,$
          where $u(V_0)$ is the probability that the impurity shifts the energy level at the site of the impurity from occupied to unoccupied states or vice versa. 
           Thus, while the average fidelity is  finite $\langle F \rangle = 1-u $,
          the typical fidelity is vanishing, 
          $\exp \langle \ln F \rangle =
          \exp ( -\infty u +  0 (1 -u )   ) =0.$  
           This statistical  mechanism for the   reduction of  the typical fidelity is 
            thereby completely different from the mechanism for the Anderson orthogonality catastrophe, where it is the coupling to a continuum of states in a metal which leads to  the power law suppression of the fidelity.

 \begin{figure*}
\centering
    \includegraphics[width=1.0\textwidth]{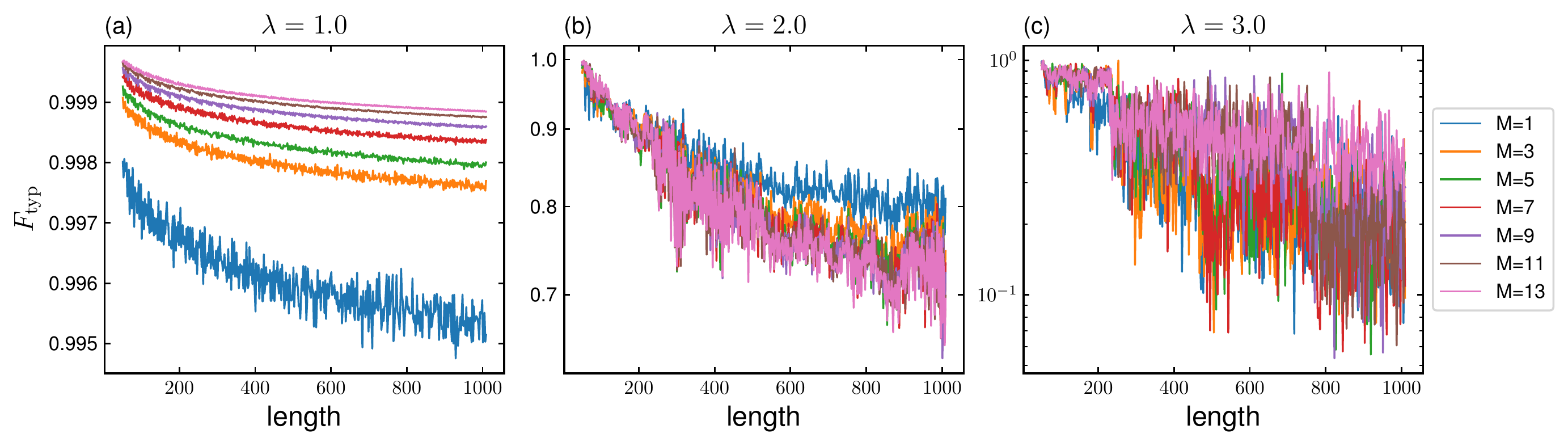}
 \caption{ Typical fidelity  at filling $n=0.309$ as function of  length $L$  (a)
  for the metallic phase $\lambda=1$, (b)
  for the critical phase $\lambda=2$, and (c) for the insulator phase $\lambda=3$.
  Results for different extensions of the impurity 
   $M$ are shown in color at fixed strength  of the  impurity $V_0=0.1J,$ sampled over $1000$ realizations. }
\label{single_typ}
\end{figure*}

       \begin{figure}
    \includegraphics[width=0.5\textwidth]{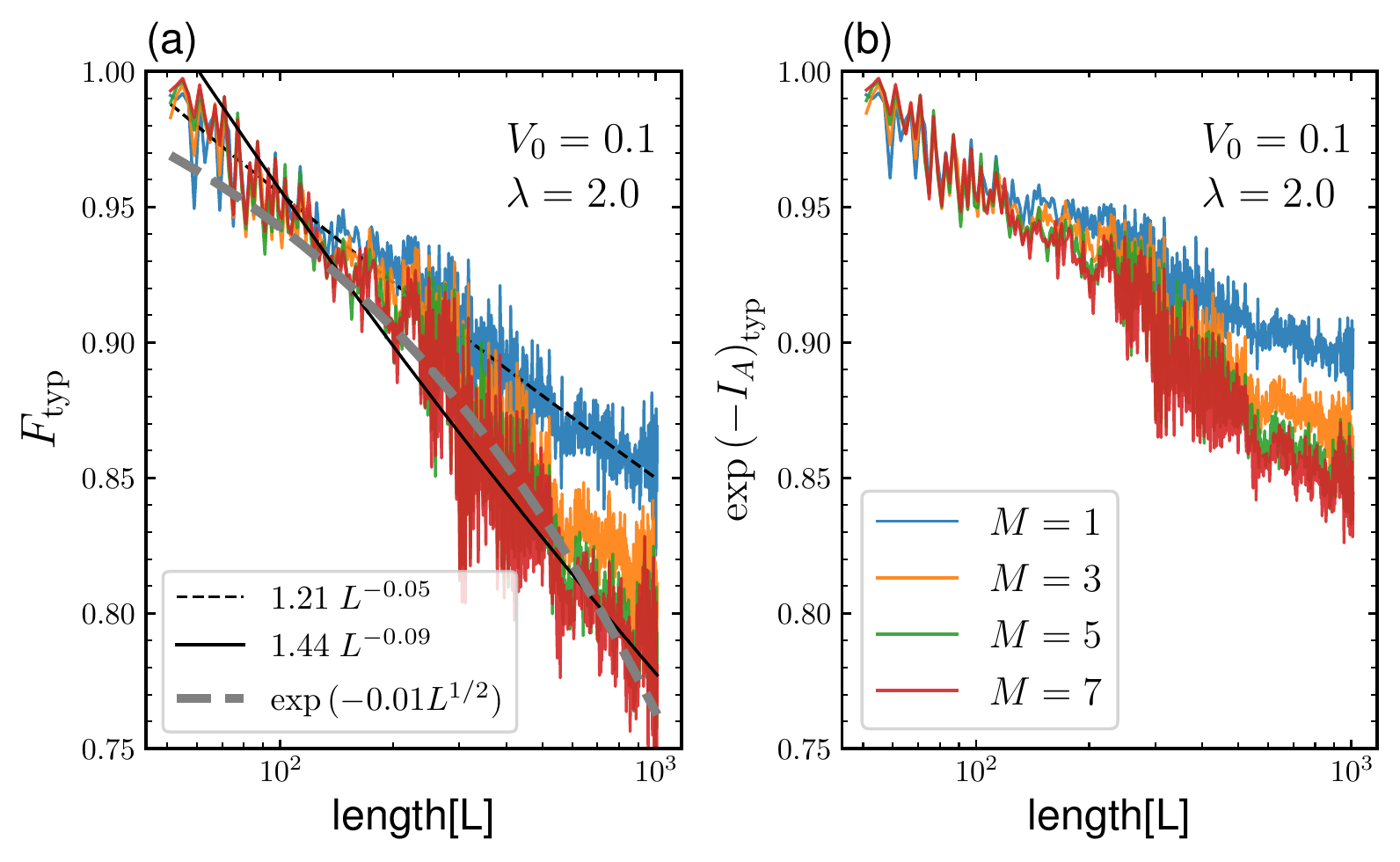}
 \caption{ (a) and (b)  show typical Fidelity and Anderson integral, respectively, as function of length $L$ for different numbers of impurity sites $M$ for a fixed impurity strength $V_0=0.1$ for the AA model at the critical point $\lambda=2.0$ and   averaged over $1000$ sample realizations.}
\label{fig10}
\end{figure} 

  \begin{figure}
\centering
    \includegraphics[width=0.5\textwidth]{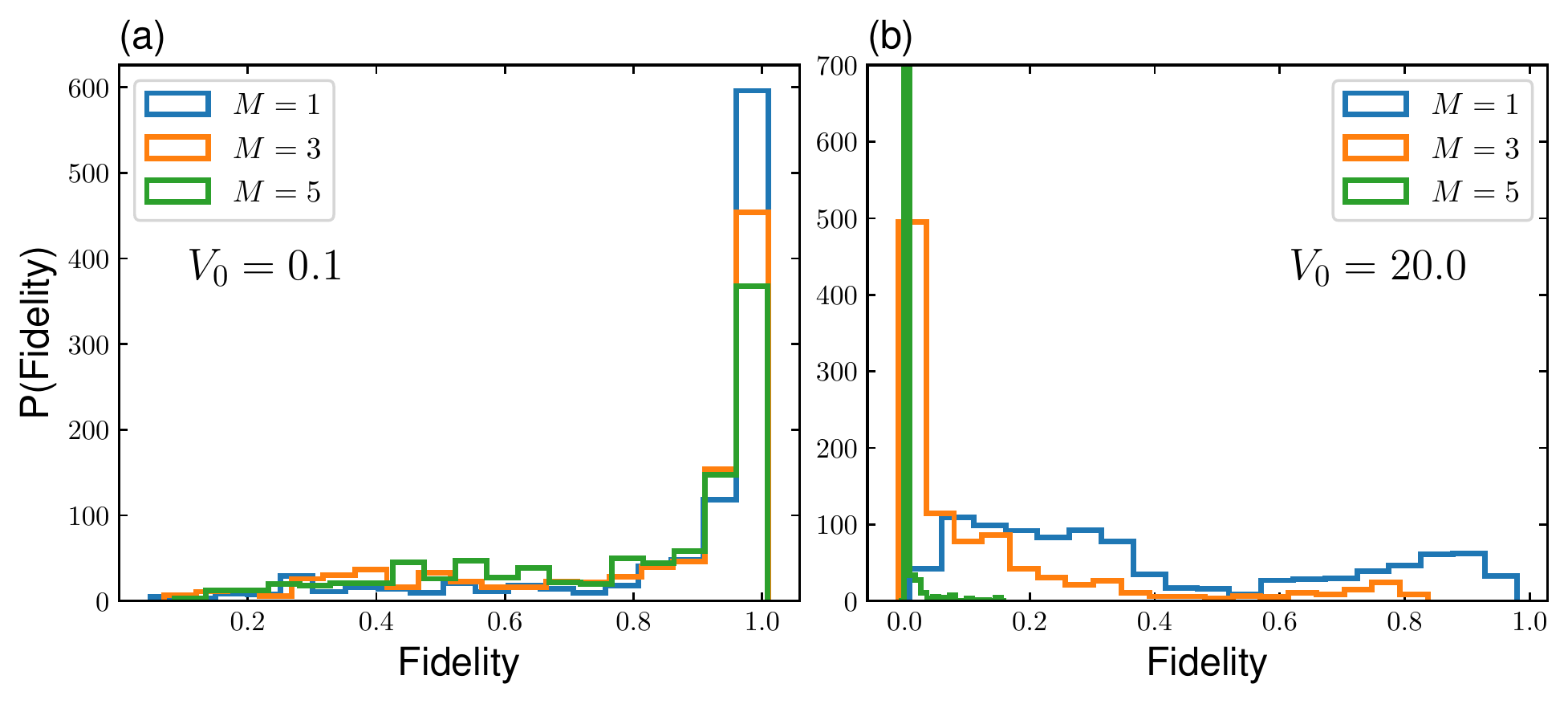}
\caption{The distribution of the fidelity $F$ in the critical phase $\lambda=2.0$
for impurity strength (a) $V_0 = 0.1$
and (b)  $V_0 = 20, $ averaged over $1000$
 realizations, for systems size $L = 1024$.}
\label{dis_lambda2_V020}
\end{figure}

 \begin{figure*}
\centering
\includegraphics[width=1.0\textwidth]{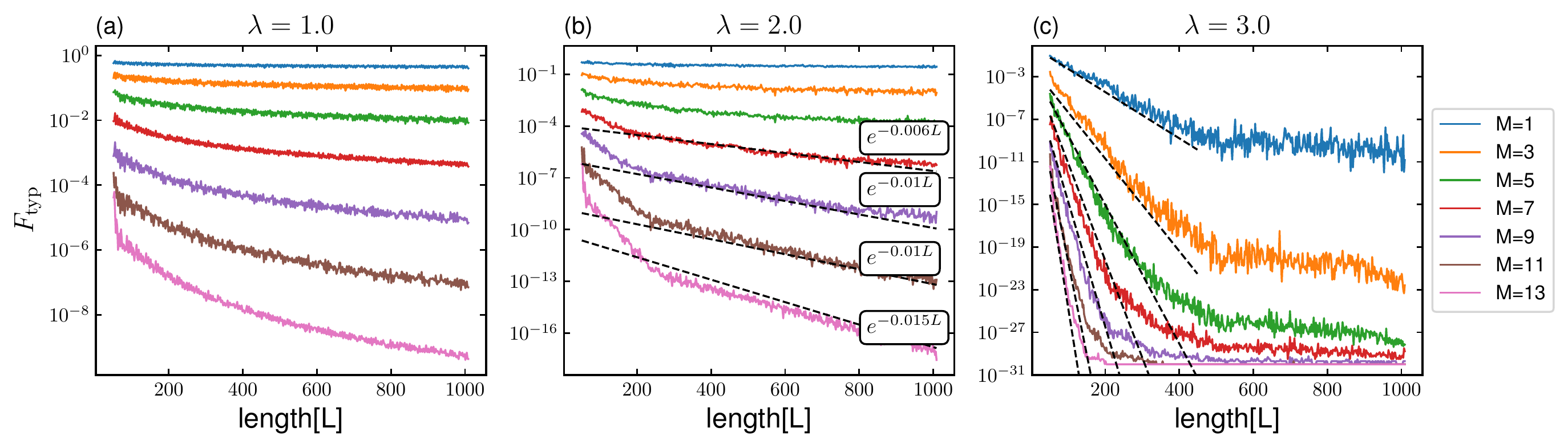}
 \caption{Typical fidelity  at filling $n=0.309$ as function of lattice length. Panels (a), (b), and (c) show results for different values of $\lambda$ as the model is in the extended, critical, and localized phases, respectively. In each panel, different data are shown in the color corresponding to the number of impurity sites $M$, while the amplitude of impurity is fixed at $V_0=20J,$ sampled over $1000$ realizations. The dashed lines are the linear fit corresponding to  exponential decay.}
\label{multi_typ_v020}
\end{figure*}

            In Fig. \ref{distribution}  the distribution of the fidelity is shown in 
             the metallic, the critical and the insulator phase for three different impurity strengths. Indeed, in the insulator phase the distribution is bimodal, and the peak around zero fidelity is increasing with impurity strength $V_0,$ approaching a Bernoulli distribution, with only small weight at intermediate values of the fidelity. 
              In contrast, in the critical phase the distribution of $F$ is very wide spreading over all values of $F,$ where the  weight of  small fidelity increases with $V_0$. The distribution of $F$ in the metal phase on the other hand has a finite width, is bimodal, and becomes shifted to smaller $F$ as $V_0$ is increased.

          Having understood the distribution of $F$, let us next try to explain the exponential suppression of the fidelity with system size $L$ in the insulator phase. As  the filling factor $n= N/L$ is fixed as the system size $L$ is increased, 
           the number of occupied levels $N$ increases.  However, in the strongly localised regime, the probability that the single state at the site of the impurity is shifted from occupied to unoccupied levels of vice versa, 
          the probability $u$,  does not change with $L,$ since it is only a function of the impurity strength $V_0,$ whether the energy level shift is sufficiently strong and the typical fidelity remains zero for all sizes. Thus, in the limit of strong single site localisation, the typical fidelity would be zero for all system sizes $L \gg 1$.

           When the  localization is not as strong, however,   each localised state  is extended  over several sites,  within the range of a  localisation length $\xi.$
            Thus, an impurity located within this range  may mix the localised state with a  finite number of other  states. 
             According to the Anderson mechanism, that would yield a finite fidelity on average, since an impurity can only be coupled to a finite number of states, which does not change as the system size increases. This is the reason that in Ref.  \onlinecite{Kettemann2016} a finite typical fidelity, independent of the system size has been found analytically. 
             However, due to the statistical mechanism, which was not considered in Ref.  \onlinecite{Kettemann2016}, the impurity may shift an occupied state up in energy so that it becomes unoccupied, or vice versa. Then, 
             another single particle state  becomes occupied  which may be (almost) orthogonal to the previously occupied state without the impurity.  As the system size increases beyond a typical localisation length $\xi,$ the  fidelity is decaying exponentially due to this statistical mechanism. As there is typically  an exponentially small but finite hybridisation matrix elements between all sites, in reality  the  impurity may couple to a larger amount of states even though with exponentially small amplitude. This might explain that 
            the typical fidelity  seems to saturate to a very small but finite value at large system size $L$ in  Fig. \ref{f_typVd20_m1}.

\section{Fidelity with Extended Impurity - Critical Exponential AOC} \label{EI}

 Next, we explore  how the fidelity depends on   the extension of the impurity at fixed total strength $V_0,$
as defined by the impurity Hamiltonian  Eq. (\ref{imp}). Clearly,  in the limit when it extends over the whole system $M=L,$ the Eigenstates 
 are not changed, and only the total energy is  shifted by $V_0/L,$ so that the fidelity is equal to one. Thus, one may expect that the fidelity increases as the 
 the extension of the impurity  $M$ is increased at fixed total strength $V_0.$
 In fact, this is what happens for a periodic 1D 
 tight binding model for a weak impurity potential $V_0 = 0.1$, as seen in Fig. \ref{tighBindingFidelity} in Appendix C, where the 
 typical value of the fidelity 
  and the upper bound $\exp (-I_A)$
are plotted  as function of chain length $L$ for different impurity extensions $M.$
 The typical fidelity increases with $M,$
 decaying more slowly with a power law of $L$, the larger $M$ is.  
 Similarly, in the metallic phase of the AA model for a weak impurity potential $V_0 = 0.1 J,$,  the typical and average fidelity become larger and decay more slowly with $L$
 as $M$ is increased, as seen in Fig. \ref{single_typ} (a), and in appendix D Fig.
 \ref{single_ave} (a), respectively,  
 in accordance with the expectation formulated above.
  Averaging Eq. (\ref{ia})
  over the phase $\phi$
  we get the 
  average Anderson integral for the extended impurity as 
    \begin{eqnarray} \label{iaaverage}
I_A  =
       \frac{ V_0^2}{2 M^2}   \sum_{n=1}^N \sum_{n'>N} \sum_{i,j \in S_M } \langle
   \frac{  \psi_{n i}^* \psi_{n' i}  \psi_{n' j}^* \psi_{n j} }{(E_{n'}-E_n )^2} \rangle_{\phi}, 
  \end{eqnarray}
 As the phase difference of the wave function amplitude between differen sites varies with $\phi$, averaging over the phase $\phi$
  gives $\langle\psi_{n i}  \psi_{n j}^* \rangle_{\phi} 
  \approx \delta_{ij} |\psi_{n i}|^2$, so that we find in  the metallic regime, where $|\psi_{n i}|^2 \sim L^{-1}$
 that   
  $I_A = 1/(2 M) \rho_0^2 V_0^2 \ln N,$
  decaying with $M$,
  resulting in an increased  fidelity with larger $M$,
  in qualitative agreement with the numerical results for the typical fidelity in the metallic regime, Fig. \ref{single_typ} (a).
 
 \begin{figure}
\centering
\includegraphics[width=0.5\textwidth]{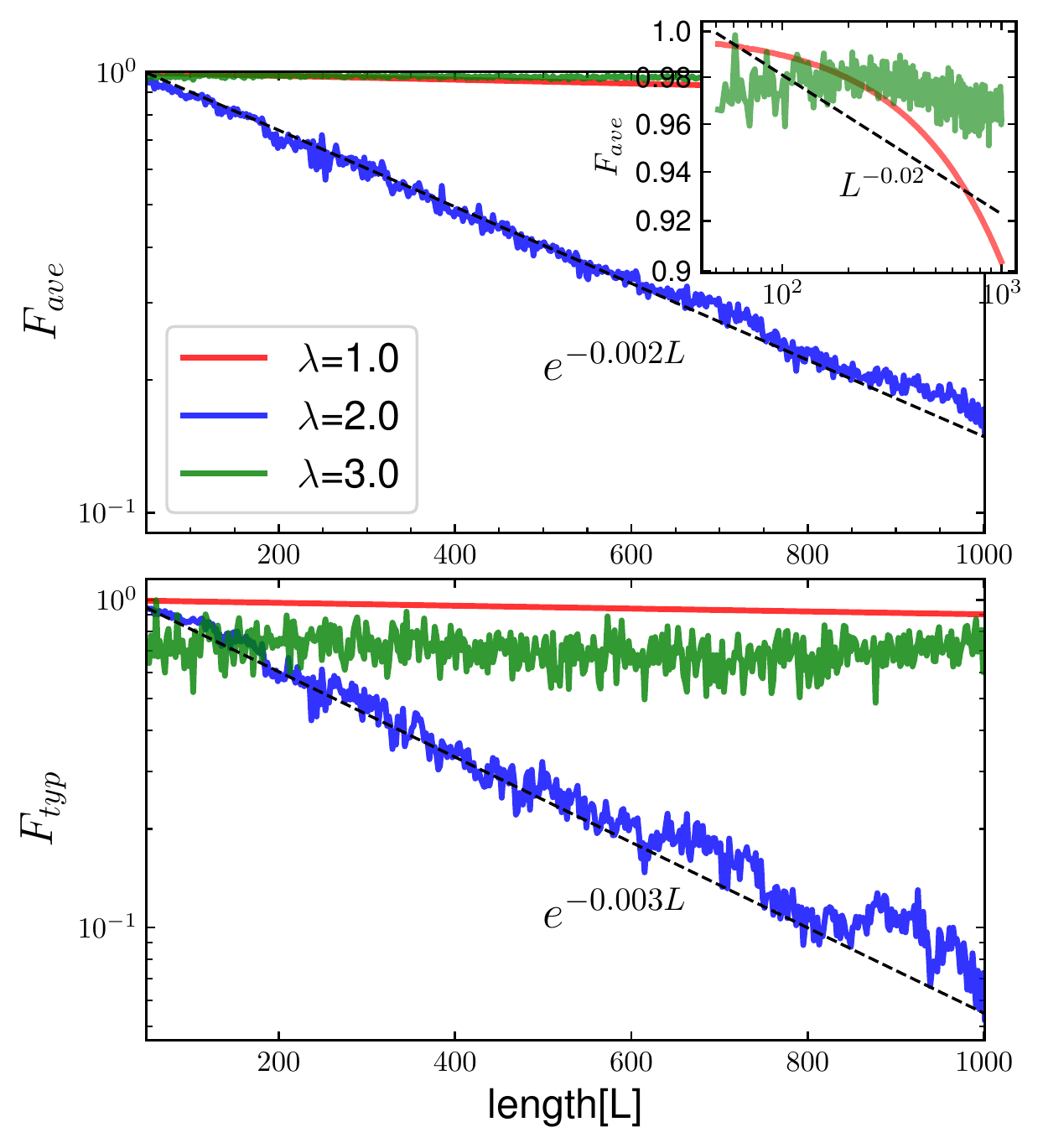}
 \caption{Average (top) and typical (bottom) fidelity after a parametric perturbation in Eq.(\ref{eqEAA}) with change $\delta\lambda=0.1$,
 for  the metallic phase  $\lambda=1$, 
 critical phase $\lambda=2$,
 and insulator phase $\lambda=3$ in solid-colored lines. The dashed-black line are the  fitted curves as indicated. Filling factor fixed at $n=0.309$, and averaged over how 1000 samples.}
\label{parametricQuench}
\end{figure}

\begin{figure*}
\centering
\includegraphics[width=1.0\textwidth]{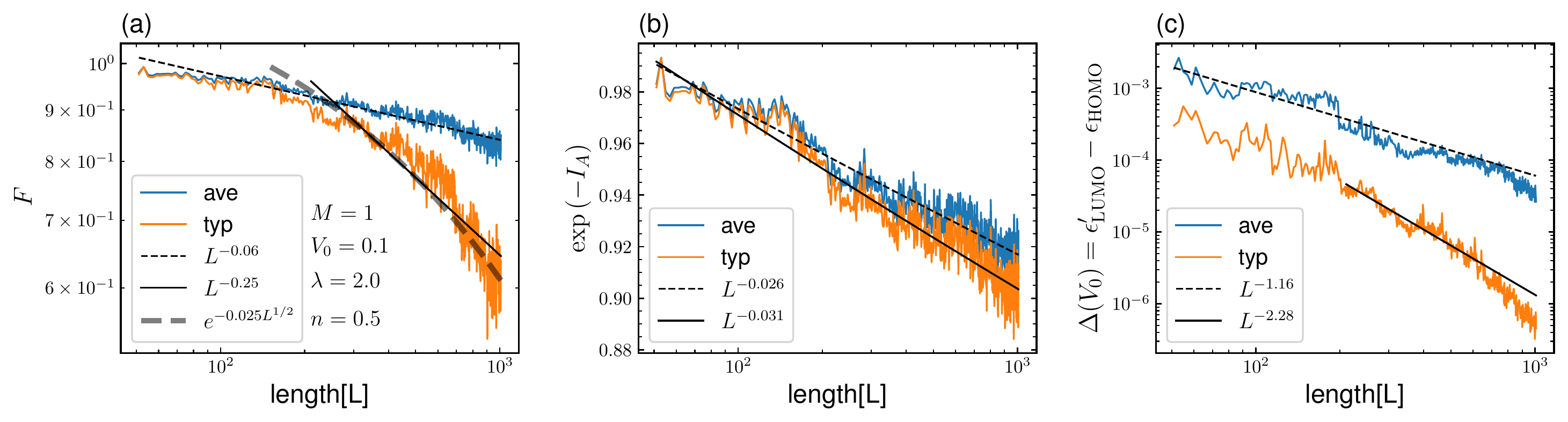}
 \caption{
For a  single  site impurity $M=1$  in the  extended AA-model at the mobility edge
for $\lambda_c=2.0$, corresponding to half filling  $n=0.5$
 (a) average and typical  fidelity  as function of  length $L$  with a single impurity of strength  $V_0=0.1$.  The dashed and solid  lines are  fits to  power laws,  as given in the legend. (b) average and typical value of  $\exp (-I_A)$, where $I_A$ is the  Anderson integral  Eq.~(\ref{ia}). (c)
average and typical value of 
the gap $\Delta (V_0)$, Eq.  (\ref{gapv0}) as function of system size $L$. The dashed and solid  lines are  fits to  power laws,  as given in the legend. 
All results are   averaged over $1000$ sample realizations.}
\label{EAA_single}
\end{figure*}

\begin{figure}
\centering
\includegraphics[width=0.45\textwidth]{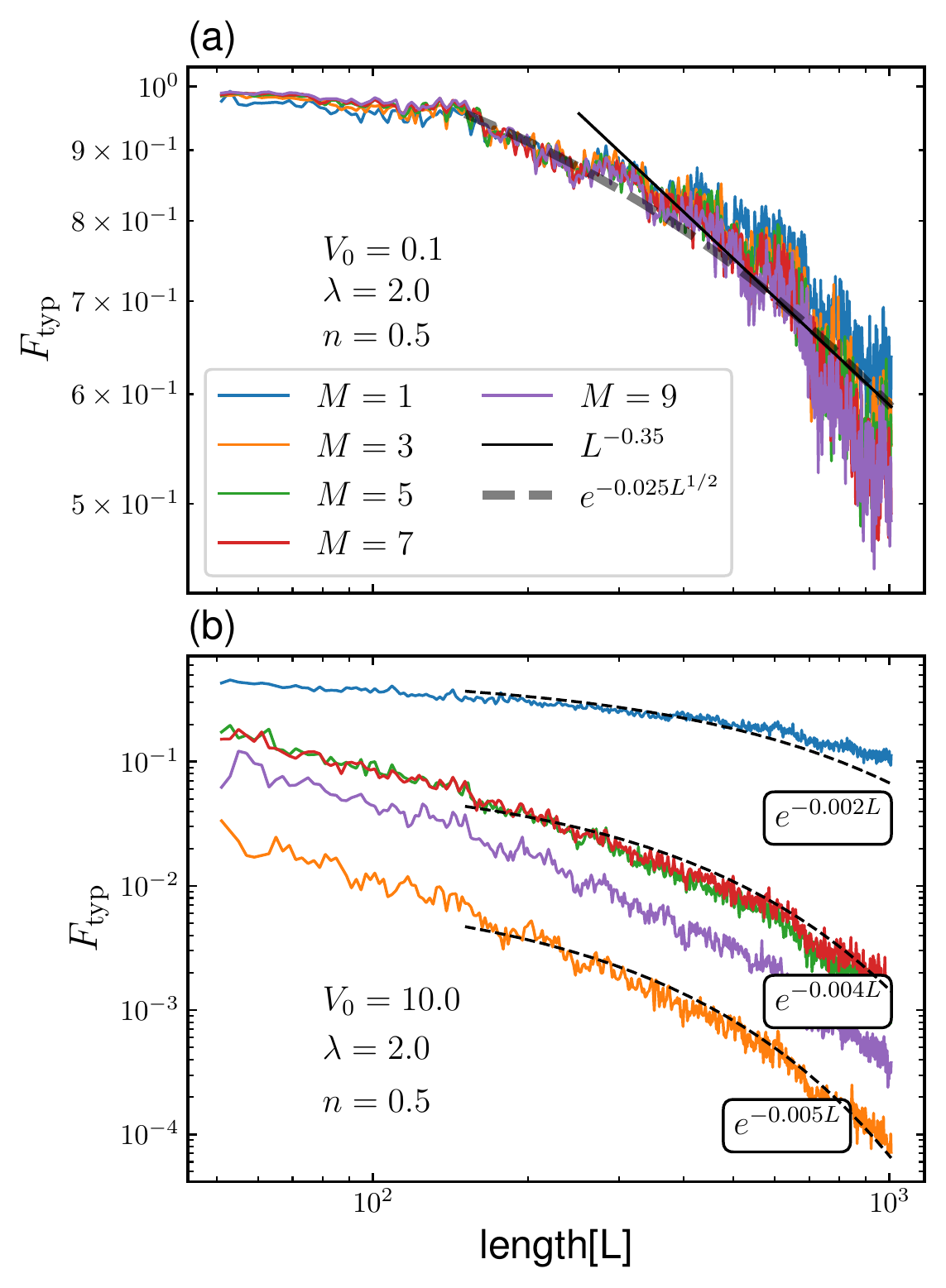}
 \caption{Typical fidelity for an  extended impurity  for 
  the extended AA-model
   at the mobility edge
for $\lambda_c=2.0$, corresponding to half filling  $n=0.5$,
  versus  length $L$ for different extension $M$. Impurity strength  (a) $V_0=0.1$, and (b) $V_0=10$.  All data   averaged over $1000$ sample realizations. Fitted curves are shown in solid or dashed black lines}
\label{EAA_multi}
\end{figure}

 In stark contrast to this,  we find 
 that in  the critical phase of  the AA-model
  the fidelity 
   is diminished more strongly  with increasing extension $M$  of the impurity, 
    as seen in Fig. \ref{single_typ} (b)
    where  the  typical fidelity is plotted, 
     as well as 
    in the Appendix C Fig. \ref{single_ave} (b),
     where 
    average  fidelity is plotted,
    as function of length $L$ for different extensions  $M$ for a  fixed, weak impurity strength $V_0=0.1$ at the critical point $\lambda=2.0,$  averaged over $1000$ sample realizations.
      Results  at the critical point 
      $\lambda=2.0,$
      are  replotted in  a semi-logarithmic plot in 
     Fig. \ref{fig10} (a)   for the  typical Fidelity and 
     in 
     Fig. \ref{fig10} (b) for the   
     Anderson integral
     as function of length $L$ for different  impurity extension $M$ for a fixed weak impurity strength $V_0=0.1$,  averaged over $1000$ sample realizations. Fits to power law and exponential dependence on $L$ are
      plotted as indicated. For the largest extension $M=7$ an exponential decay cannot be excluded. 
Thus, we may  recover the analytically predicted exponential AOC in the critical phase\cite{Kettemann2016}, albeit only for an extended impurity.  A possible explanation is that the magnitude of multipoint correlations, which we found to be responsible for masking the critical two point correlations, may become diminished,  
for the extended impurity, so that the critical enhancement of two point correlations dominate the typical fidelity for extended impurity, resulting in the exponential AOC in the critical phase.
    To get a better understanding 
     of the result in the critical phase
      let us look at the  distribution of the fidelity $F$ in the critical phase $\lambda=2.0$.
For impurity strength  $V_0 = 0.1$ averaged over $1000$
 realizations, for systems size $L = 1024$ the distribution  is shown in Fig.
\ref{dis_lambda2_V020} (a). The distribution 
 of the fidelity is found to be wide, as expected 
  in the critical phase\cite{Kettemann2016}. The probability that the ground state is not affected by the impurity, that the fidelity is close to one, is found to decrease with increasing $M.$

  In 
  the insulator phase $\lambda =3$ for a fixed weak impurity strength $V_0=0.1$
  the fidelity is 
  decaying  more strongly than in the other phases. 
   With increasing extension $M$  of the impurity the fidelity becomes larger, 
    as seen in Fig. \ref{single_typ} (c) and
    Fig. \ref{single_ave} (c),
    where  the  typical fidelity and
    average  fidelity are plotted, respectively,
    as function of length $L$ for different  $M$   averaged over $1000$ sample realizations. 
    The reason for that behavior might be that the effect on single site localised states is smaller for an extended impurity, thereby diminishing the  probability $u$ that an occupied state becomes shifted to unoccupied states as the impurity is turned on, enhancing thereby the fidelity, according to the theory of the statistical exponential AOC as outlined in section  
    \ref{seoc}.

       For an  extended  impurity with 
  strong amplitude $V_0 =20 J,$ 
  we find that 
  the  
 AOC is clearly exponential 
  in the critical phase 
as seen  in Fig. 
  \ref{multi_typ_v020} (b)
   for extension $M \gg 1.$
We find that the larger the extension $M$,  the  more the fidelity becomes   diminished and  the  stronger the exponential AOC becomes. This is confirmed  by the distribution of the fidelity  for such a strong impurity in the critical phase, as shown in Fig.   \ref{dis_lambda2_V020} (b).
 
In the metallic phase  a strong impurity
     $V_0 = 20 J$ is found to reduce the fidelity more strongly with increasing $M,$ but  the typical
     fidelity  Fig. 
  \ref{multi_typ_v020} (a)
     and   average fidelity  Fig. 
     \ref{multi_typ} (a)
     continue to decay with a power law in $L$ for all  $M.$ A similar behavior is found for a strong impurity in the tight binding model  as shown in Appendix C, Fig. \ref{tightv10}.
    
      In the insulator phase Fig. 
  \ref{multi_typ_v020} (c) and 
  Fig. 
     \ref{multi_ave} (c)
  shows a strong exponential AOC
  for the typical and average fidelity respectively, 
  which becomes stronger with the extension $M$ of the impurity.

\section{ Fidelity with Parametric Perturbation - Parametric Exponential AOC} \label{EI}
  
  The concept of fidelity has  been generalized
    to  parametric perturbations of a quantum system. It   
     have  been  successfully used to 
     characterize quantum phase transitions \cite{venuti}.
     Therefore, let us next study 
     a perturbation which shifts  
      the parameter $\lambda$ by a
       small amount $\delta \lambda$
  \begin{equation}
H_{Pert} =\delta \lambda \sum_{i=1}^L \cos (2 \pi Q i +
\phi)  c_i^+ c_{i}.
\label{eqAA_per}
\end{equation}

  The effect of such  a parametric quench  has been  recently  studied in  the AA-model in Ref \cite{Bobo2019} 
  by calculating the so called fidelity susceptibility
  $\chi_F(\lambda) = \lim_{\delta\lambda\rightarrow0}
  -2\log{F}/\delta\lambda^2$.
  We note that, an upper bound for
  the fidelity susceptibility 
  $\chi_F(\lambda)$ is given 
   by  the Anderson Integral, 
   \begin{eqnarray}
   \chi_F(\lambda) 
   &\le & \lim_{\delta\lambda\rightarrow0} 2 I_A/\delta\lambda^2 \nonumber \\
   &= &\lim_{\delta\lambda\rightarrow0}\frac{1}{\delta\lambda^2} 
   \sum_{n=1}^N \sum_{n'>N}
   \frac{ |  \langle n | H_{Pert} | n' \rangle|^2}{(E_{n'}-E_n )^2}\end{eqnarray}
   \begin{equation} \nonumber
   = 
   \sum_{n=1}^N \sum_{l>N} \sum_{i,j}
   \frac{\cos (2 \pi Q i +
\phi)\cos (2 \pi Q j +
\phi)  \psi_{ni}^* \psi_{nj} \psi_{li} \psi_{lj}^*}{(E_{l}-E_n )^2},
   \end{equation}
   where the indices $n,l$ denote  the unperturbed Eigenstates. 
   
   In the critical phase, noting that 
    local intensities are power law correlated in 
     energy, the dominating contributions come
      from the terms at the same locations
       $i=j.$
        Thus we find
        \begin{equation}
         \chi_F(\lambda) \le 
          \sum_{n=1}^N \sum_{l>N} \sum_{i}
   \frac{\cos (2 \pi Q i +
\phi)^2  |\psi_{ni}|^2 | \psi_{li}|^2 }{(E_{l}-E_n )^2}.
   \end{equation}
   Averaging over the phase $\phi$ we thereby find
   approximating $\rho(E) \approx \rho_0$,
   and using $\Delta = \Delta_0 L^{-z}$
    \begin{equation} \label{chiana}
         \chi_F(\lambda) \le 
        \frac{1}{2}  \frac{\rho_0^2}{\gamma (1+\gamma)}\left( \frac{D}{\Delta_0}  L^z \right)^{\gamma} \sim L^{z \gamma}
   \end{equation}
   where for the AA model in the critical phase $\lambda_c=2,$  $\gamma=1/2$.

  Fig.~\ref{parametricQuench} shows the results for average (upper figure) and typical (lower figure)  fidelity of a  parametric  perturbation with $\delta \lambda =0.1$.  As shown,  in the metallic phase $\lambda<2$ we find a power-law decay in the average and typical fidelity, both  with the same  power, scaling with $L$ as $L^{-0.02}$.
  
  At the critical point $\lambda_c=2$ the average and typical fidelity 
  are clearly found to 
  decay exponentially.
  The  average fidelity is found to decay as $F_{ave} \sim
  e^{-0.002 L}$ and the typical as $F_{typ} \sim e^{-0.003 L}$.
  Thus, this gives for the typical fidelity susceptibility 
  $\chi_F \approx 0.6 L,$
  in good agreement with the analytical upper bound Eq. (\ref{chiana}), which 
  gives with $\gamma =1/2,$
   $\chi_F < 2/3 L^{z/2},$
   where $z$ is the dynamical exponent, which we found numerically to be close to $z \approx 2.$
  
   Using another approach near the quantum critical point $\lambda_c=2$, it was  recently argued that the  fidelity susceptibility scales with system size as $\chi_F(\lambda_c)\sim N^{2/\nu}$\cite{Bobo2019}, where $\nu$ is  the correlation length critical exponent, given by
 $\nu \approx 0.89$ according to Ref.  \onlinecite{Thakurati2012}
  and by 
 $\nu \approx 0.95$ according to  Ref \onlinecite{Cookmeyer2020}, which is also in some agreement with our numerical results.
  
 In the localized phase $\lambda =3$  both the average 
 and typical 
 fidelity show a very weak and strongly fluctuating  dependence on $L$.

\section{ Fidelity in the Extended AA Model} \label{EAA}

The extended AA (EAA)-model with Hamiltonian Eq. (\ref{eqEAA}) for $b>0$ has a mobility edge, as seen in  Fig. \ref{ipr} (right) where the
 inverse participation ratio is plotted  as function  of  energy   and parameter $\lambda$ for the  EAA-model where a mobility edge (black solid line) separates the extended phase  with ${\rm IPR}\rightarrow0$ from the localized 
 ${\rm IPR} \rightarrow a/\xi,$ where $xi$ is the localization length and $a$ the lattice spacing.
Let us therefore explore whether 
there is a critical exponential AOC 
 at the mobility edge, as found analytically  in Ref.\cite{Kettemann2016}  at a mobility edge. 
 
 First we address the case with a single site  impurity $M=1$  with weak potential $V_0=0.1$. In our calculation we set $b=0.2$ and  $\lambda=2.0J$. We choose half filling  $n= N/L =0.5$ 
  so that the Fermi energy is 
 at the mobility edge  $E_{\rm mb}=0,$ see Fig.~\ref{ipr}. 
In Fig.\ref{EAA_single} 
we plot both  the average and typical fidelity. The fit 
 with a power law in system size $L$
 is good for the average fidelity. The typical fidelity shows a much  smaller value for all system sizes with a stongrt  decay with $L$.
 The decay becomes stronger at larger $L$ possibly indicating  an exponential AOC. 
 
 Compared with the  single site impurity 
 with weak potential $V_0 = 0.1 J$ in the 
 AA model,  $b=0$,
  in the critical phase $\lambda_c=2$, shown in 
  Fig.~\ref{fig:singlesite}) (a), 
   the average fidelity is of similar magnitude as in the EAA model at the mobility edge Fig.\ref{EAA_single},  while  the typical 
   fidelity is smaller and decays  faster in the EAA model at the mobility edge Fig.\ref{EAA_single}. 
 
 We depicted the average and typical value of  $\exp(-I_A)$
in  Fig.\ref{EAA_single} (b) 
 and  find that it  gives   as expected  an upper bound for the fidelity for the whole range of systems sizes explored. But we note that the difference between the average and typical value is not as profound as for the fidelity.  

We also plot the average and typical gap, the difference between HOMO and LUMO energies  $\Delta(V_0)$, Eq.(\ref{gapv0}) 
for the extended AA model at the mobility edge in 
Fig. \ref{EAA_single} (c).
For both average and typical, we find  a power law decay  with $L$ with dynamical exponent $z>1$. Interestingly, the  decay of the gap is stronger  for  the typical  than the average gap, which is in contrast to the result for the gap  Eq. (\ref{gapv0})  in the critical regime in the AA model, where the average and typical gap showed a similar magnitude and decay, see Fig.  \ref{gapsingle}.

Finally, let us  consider
 the fidelity in the EAA model at its mobility edge  with  an  impurity extended over $M$ sites. 
 In Figs.~\ref{EAA_multi} (a) and (b) we present the typical fidelity for two impurity strengths $V_0=0.1$ and $V_0=10$, respectively. For the weak impurity $V_0=0.1$ the typical fidelity decays with a  power law  decaying with system sizes to smaller values the larger the extension of the impurity $M$ is. Thus, this is a similar behavior as  we have observed for a weak impurity in the critical AA-model  in  Fig.\ref{single_typ}.   
 For the strong impurity we observe  in Fig.~\ref{EAA_multi}  (b) a smaller typical fidelity decaying exponentially with system size , similarly as  
for a strong  impurity in the critical AA-model  in  Fig.\ref{multi_typ_v020} (b).

\section{Conclusion} \label{C}
We  presented
evidence for the  exponential 
 orthogonality catastrophe
in the    (extended)
  Aubry-Andr\'e (AA)-Model with an added potential impurity, as function of the size of 
        that impurity.
   While  we  not find 
   a predicted exponential AOC in the critical regime
   for a  weak single site impurity, but rather find  that the fidelity decays with a power law,  in the  critical phase. Even though it is found to be  smaller and decays  faster than in the metallic phase, it does not decay exponentially as predicted.   For an extended impurity, however, we find indications of  an 
     exponential AOC at  the quantum critical point of the AA model and at the mobility edge of the extended AA model and suggest an explanation for this finding.  By reexamination of the analytical derivation we identify nonperturbative corrections due to the impurity potential and  multipoint correlations among wave functions as possible causes for the absence of the exponential AOC in the critical phase. 
     
     We find a different kind of 
    exponential AOC  in the insulator phase
    for which we give a
    statistical explanation, 
    similar to that it was  given in  \cite{Khemani} for an adiabatic perturbation in an insulator phase,  
    a mechanism which is  
     profoundly 
    different from the AOC in metals, where it is the coupling to a continuum of states which yields to the power law suppresion of the fidelity.

     Furthermore we consider a  
     parametric perturbation to the AA model, and find an exponential AOC numerically, in agreement with an analytical derivation 
      which we
     provide here.
     
      It has been suggested that  the orthogonality catastrophe  
       can be studied in ensembles of ultracold atoms in a
      controlled way\cite{demler}.
      Indeed, since
      the 
       extended AA model was introduced and suggested to   be experimentally realised   in atomic optical lattices and photonic wave guides\cite{Ganeshan2015}, it was 
       recently realized in 
        synthetic lattices of laser-coupled atomic momentum modes, and demonstrated to have  a   mobility edge\cite{An2021}.
       We therefore hope that our analysis 
        will  provide guidance for  the experimental study of the fidelity and the AOC in these systems. 
        Furthermore, this opens new pathways for 
      the study of nonequilibrium quantum dynamics. We note that our results can be extended to  interacting disordered fermion systems, as multifractality exists even in strongly interacting disordered systems\cite{Amini2014}.

{\bf Acknowledgement.}           
We  gratefully acknowledge  the support from Deutsche Forschungsgemeinschaft (DFG) KE-807/22-1.   We thank  Eugene Demler for stimulating discussions, initiating this study and 
 Keith Slevin for stimulating discussions
  and useful comments.

\subsection{Appendix}
  When the single particle states of a Fermi system are $ |n \rangle = c^+_n |0 \rangle$, 
   the ground state of $N$ fermions is given by
    $| \psi  \rangle = \prod_{n=1}^N c_n^+ |0 \rangle.$
     Adding an impurity the single particle states are 
     changed to $ |n' \rangle = c^+_{n'} |0 \rangle$, 
   so that the ground state becomes
    $| \psi'  \rangle = \prod_{n'=1}^N c_{n '}^+ |0 \rangle.$
   The fidelity is given by the absolute value of the scalar product, 
    $F= |\langle \psi | \psi' \rangle | = | \langle 0 | \prod_{n=1}^N  c_{n }
     \prod_{n '=1}^N c_{n '}^+ |0 \rangle|.$
     Defining the scalar product of single particle states of the pure system and the system with 
      perturbation, $A_{n n'} = \langle n | n' \rangle$, and 
     applying  the anticommutation relations for $c_n, c^+_n$,  
     we can write the fidelity as 
     \begin{equation} \label{exactfidelity}
      F = |{\rm det}_{n,n ' \le N} A|.
      \end{equation}
      $A_{n n'}$ is for fixed $n$ a normalized vector with  $ \sum_{n'} |A_{n n'}|^2=1,$ 
     However, since the summation in the fidelity
     $F$ is restricted, and  
     $ |n' \rangle = \sum_{n \le N} A_{n n'} | n \rangle 
     + \sum_{n >N} A_{n n'} |n \rangle,$
      only those vector components with $n \le N$ contribute to the fidelity $F$.
       This means  that the determinant is taken of a square matrix with column vectors which are not normalized. However, we can 
       normalize each column vector, by multiplying it with 
        $c_{n'} = (1-  \sum_{n >N} |A_{n n'}|)^{-1/2}$ and get the identity
        \begin{equation}
        F =  \prod_{n' \le N} (1-  \sum_{n > N} |A_{n n'}|^2)^{1/2}  {\rm det} (A  \prod_{n' \le N}c_{n'}).
        \end{equation}
        Since the second factor is now a determinant with normalized column vectors, it cannot exceed one, 
         but can be samller, so that 
         $ det (A  \prod_{n' \le N}c_{n'}) <1$, and therefore
           \begin{eqnarray}
        F <  \prod_{n' \le N} (1-  \sum_{n>N} |A_{n n'}|^2)^{1/2} 
        \nonumber \\
        < \exp ( - \frac{1}{2}\sum_{n' \le N} \sum_{ n > N}   |A_{n n'}|^2). 
        \end{eqnarray}

\subsection{Appendix}

When the unperturbed system has the Hamilton operator 
 $H_0$ with Eigenstates $|n \rangle$ determined by the Schr\"odinger equation 
  $H_0 |n \rangle= E_n |n \rangle,$
   adding an  impurity with 
   Hamiltonian $H_{\rm imp}$, Eq. (\ref{imp})  with potential strength $V_0$
   changes the  Eigenstates  to $|n' \rangle$ as determined by 
     $(H_0 +  H_{\rm imp}) |n' \rangle= E_{n'} |n' \rangle.$
     Multiplying by left with $\langle n | $ we thus get  the identity 
      \begin{equation}
      \langle n | n' \rangle =  \frac{1}{E_{n'}-E_n }    \langle n | H_{\rm imp} | n' \rangle. 
      \end{equation}
      Thus for a local impurity $V = V_0 \delta ({\bf r}- {\bf x})$, we find 
       \begin{eqnarray} 
I_A &= &\frac{1}{2} \sum_{n \le N, n'> N}
|\langle n | n' \rangle|^2 \nonumber \\ &=& 
           \frac{1}{2} \sum_{n \le N, n'>N}
       \frac{1}{(E_{n'}-E_n )^2}  |  \langle n | V | n' \rangle|^2  \nonumber \\ &=& 
           \frac{ V_0^2}{2} \sum_{n \le N, n'>N}
       \frac{ | \psi_n({\bf x}) |^2 |\psi_{n'} ({\bf x})|^2}{(E_{n'}-E_n )^2} , 
      \end{eqnarray}

      where $| \psi_n({\bf x}) |^2 = | \langle n | {\bf x }\rangle|^2$,  
      $| \psi_{n'}({\bf x}) |^2 = | \langle n' | {\bf x }\rangle|^2$, is the intensity 
       with and without the additional impurity at postion ${\bf x }.$ 
       
       For a single site potential impurity at site ${\bf x }$
       with amplitude $V_0$,
        the perturbed intensity 
        $| \psi_{n'}({\bf x}) |^2$
        can be written as\cite{Economou1983}
          \begin{eqnarray}
 |\psi_{n'} ({\bf x})|^2 = \lim_{ E \rightarrow E_{n'}}(E-E_{n'}) 
           \frac{V_0 (G_E^0({\bf x},{\bf x}))^2 }{1- V_0 G_E^0({\bf x},{\bf x})}, 
      \end{eqnarray}
      where, 
      \begin{equation}
      G_E^0({\bf x},{\bf x}) = 
      \sum_l|\psi_{l} ({\bf x})|^2   
      \frac{1}{E-E_l+\i \delta}.
      \end{equation}
         Performing the limit using  de l'Hospital, one finds
          \begin{eqnarray}
 |\psi_{n'} ({\bf x})|^2 =  
           \frac{ (\sum_{l}
       \frac{ | \psi_l({\bf x}) |^2 }{E_{n'}-E_l } )^2 }{\sum_{m}
       \frac{ | \psi_m({\bf x}) |^2 }{(E_{n'}-E_m )^2}}, 
      \end{eqnarray}     
      which depends on the disorder potential only implicitly 
      through the Eigen energy of 
       the perturbed state $E_{n'}.$
      
      Keeping in all summations only 
       the largest terms, 
        which are the terms with 
       the unperturbed Eigen energy  $E_l$ closest to $E_{n'},$
         we get approximately
            \begin{eqnarray}
 |\psi_{n'} ({\bf x})|^2 =  
            | \psi_l({\bf x}) |^2|_{min_l(|E_l-E_{n'}|)} , 
      \end{eqnarray}
      and we recover the result obtained in second order perturbation theory.

   \subsection{Appendix}

 In this appendix, the tight-binding model is revisited numerically as a benchmark. We consider Hamiltonian $H=\sum_i(c_i^\dagger c_i+h.c)$ and introduce the impurity as is defined in the main text Eq.(\ref{imp}).

\begin{figure*}
\includegraphics[width=1.0\textwidth]{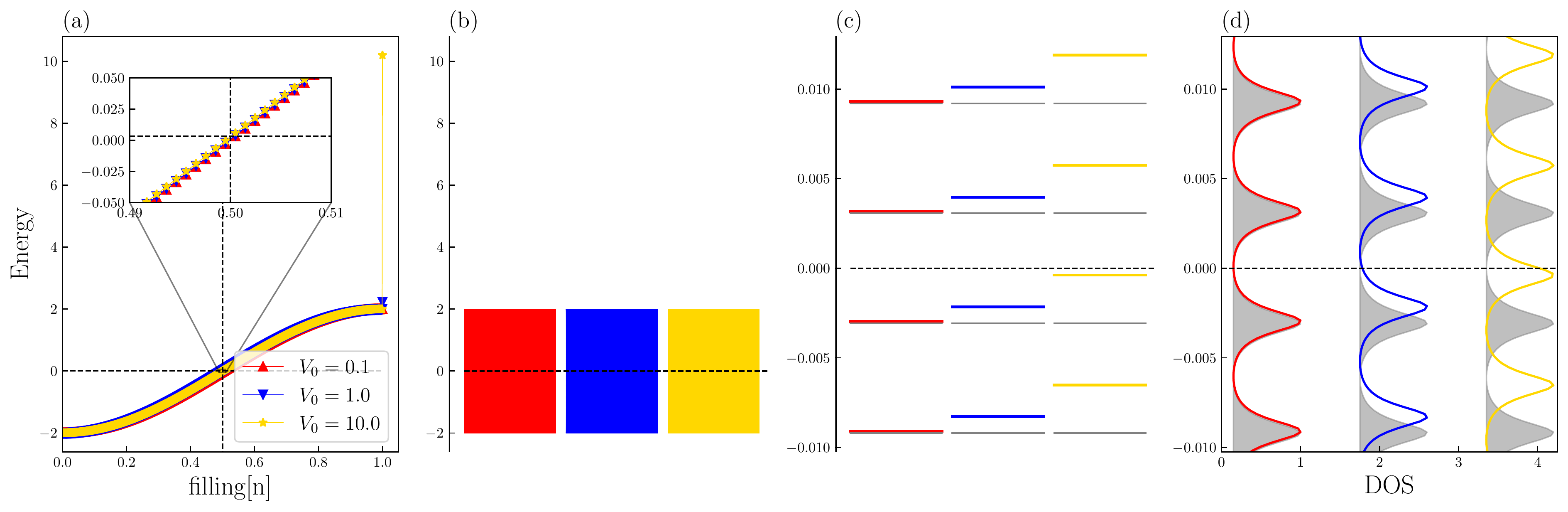}
 \caption{Energy diagram and density of state for the tight binding model with a single impurity $M=1$ randomly sitting on the chain.}
\label{tighBinding1}
\end{figure*}

 Fig. \ref{tighBinding1}-(a)  shows the energy level spectrum as function of filling factor $n$ for a single impurity $M=1$ with three different strength $V_0$ (as displayed by the colored symbols). The  dashed line indicates the filling of $n = 0.5$,  corresponding without an impurity  to the Fermi energy $E_F/J=0$. Inset shows a zoom close to the Fermi energy. Fig. \ref{tighBinding1} (b) and (c) show a full and zoomed energy level diagram, with and without impurity. The case without impurity is drawn in a grey color. It can be seen bigger impurity strength has strong shifts of the energy close the Fermi energy.  Fig. \ref{tighBinding1}  (d) shows the density of state (DOS) as a function of energy close to Fermi energy.
 
\begin{figure*}
\includegraphics[width=1.0\textwidth]{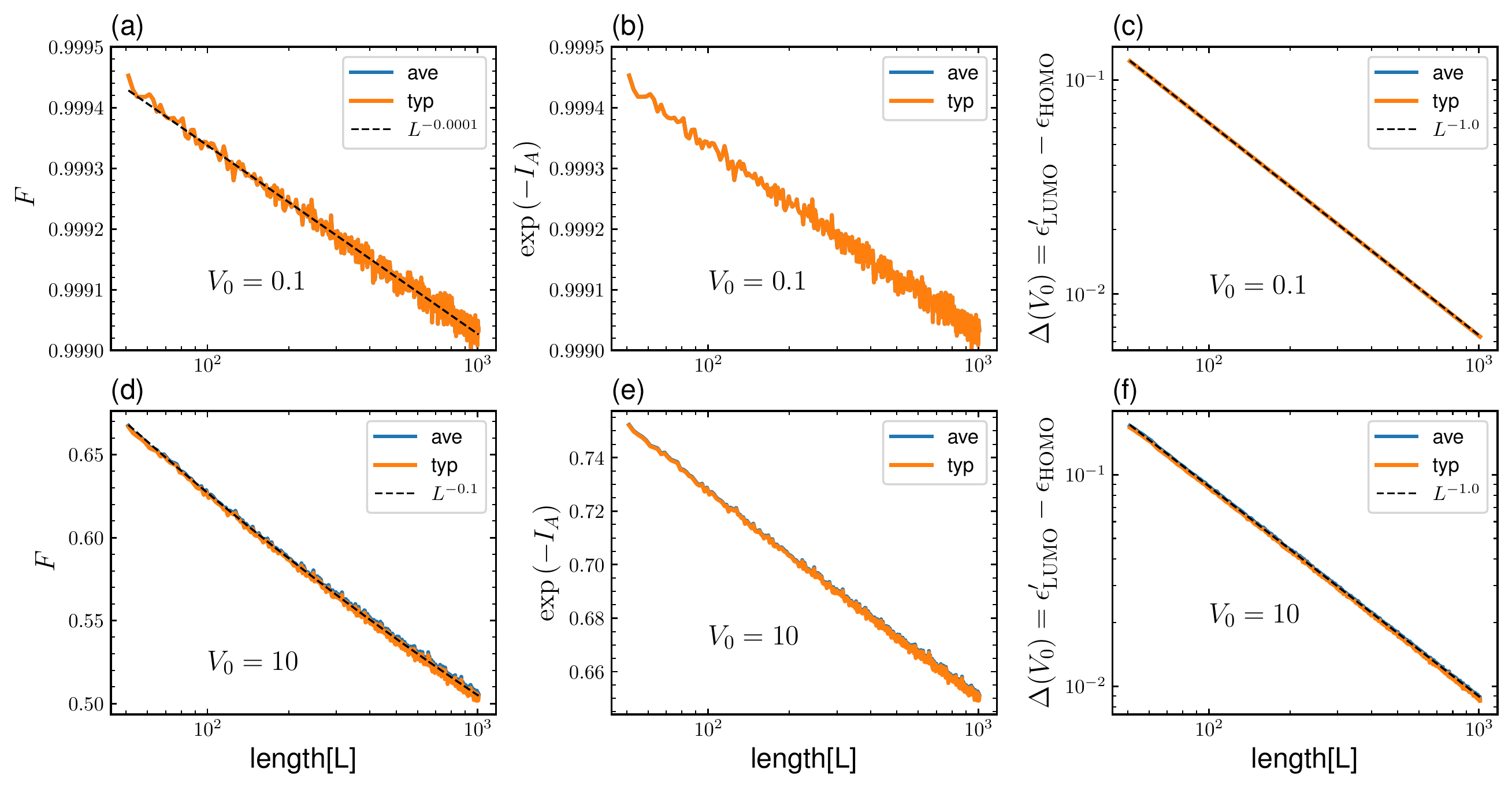}
 \caption{Average and typical value of the fidelity, Anderson integral and gap evolution as a function of chain length, for the tight-binding model with a single impurity $M=1$ randomly sites on the chain. Data averaged over $1000$ realizations. Black-dashed lines are fitted curves. }
\label{tighBinding2}
\end{figure*}

 In Fig.\ref{tighBinding2} we show the typical and average of the fidelity $F$, the Anderson integral $I_A$, and the gap $\Delta$ evolution as a function of system size $L$ for two impurity strength $V_0=0.1,~10$. In numerical calculation we considered a single impurity randomly located on the chain and averaged over the position. As can be seen the typical and average are the same as expected for the clean model. For impurity strength $V_0=0.1$, we found a small power-law decay ($L^{-0.0001}$) for both fidelity and $I_A$ as function of system size. We noticed that fidelity almost touches the $I_A$ as an upper bound limit for all the range of system size shown here.  While for the strong impurity case, $V_0=10$, fidelity decays much faster with system size ($L^{-0.1}$) and smaller than the $I_A
$ for all ranges of system size. However, in both weak and strong impurity, the gap $\Delta$  independent of the impurity strength and decays in power-law as $L^{-1}$. 

\begin{figure}
\includegraphics[width=0.5\textwidth]{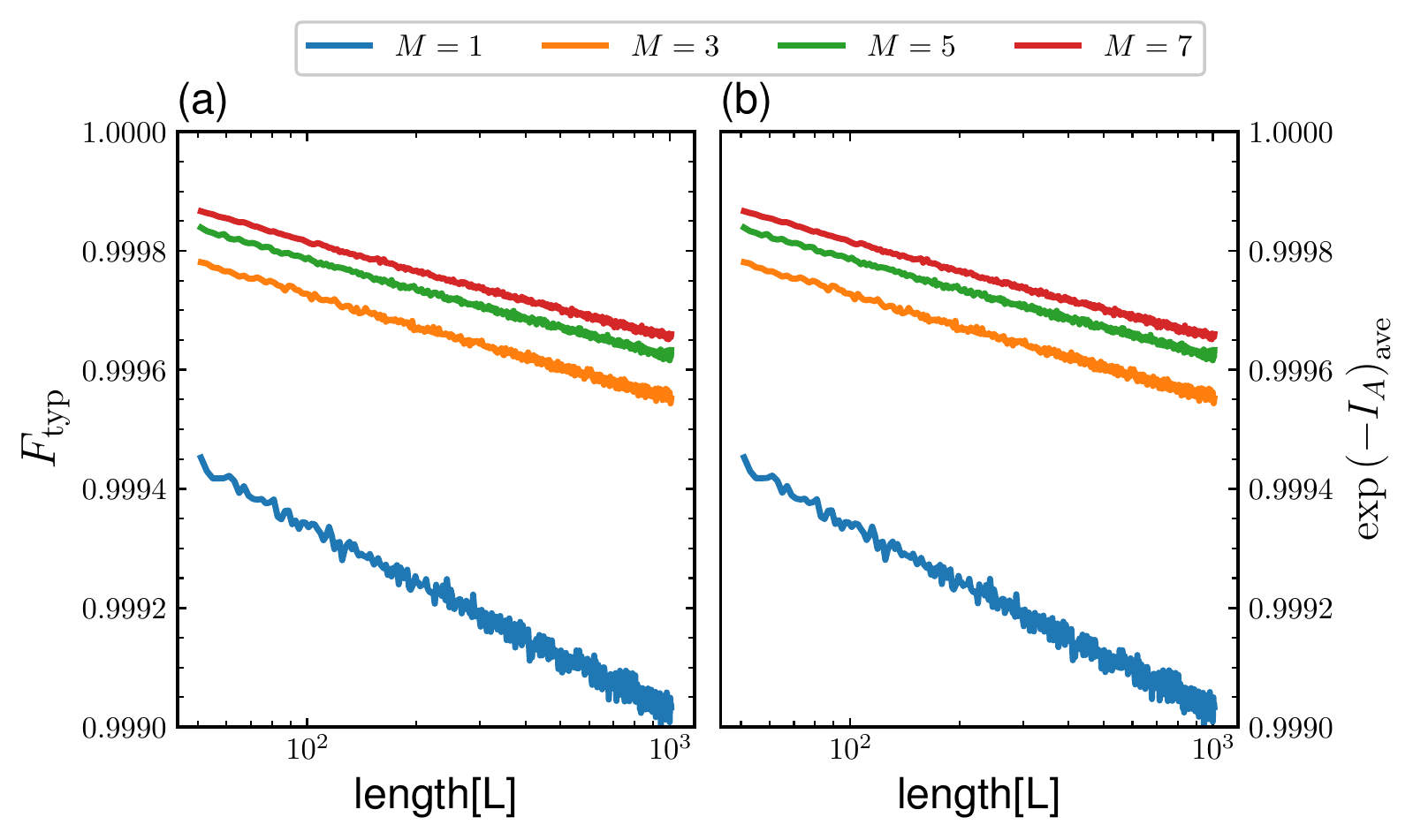}
 \caption{ Typical value of (a) the fidelity 
  and (b) the upper bound $\exp (-I_A),$ for a 1D tight binding model 
 plotted  with added  impurity 
 Eq. (\ref{imp}) as function of chain length $L$ for different impurity extensions $M$  for fixed  impurity strength $V_0= 0.1$.  Data averaged over $1000$ realizations.}
\label{tighBindingFidelity}
\label{tightv01}
\end{figure}

\begin{figure}
\includegraphics[width=0.5\textwidth]{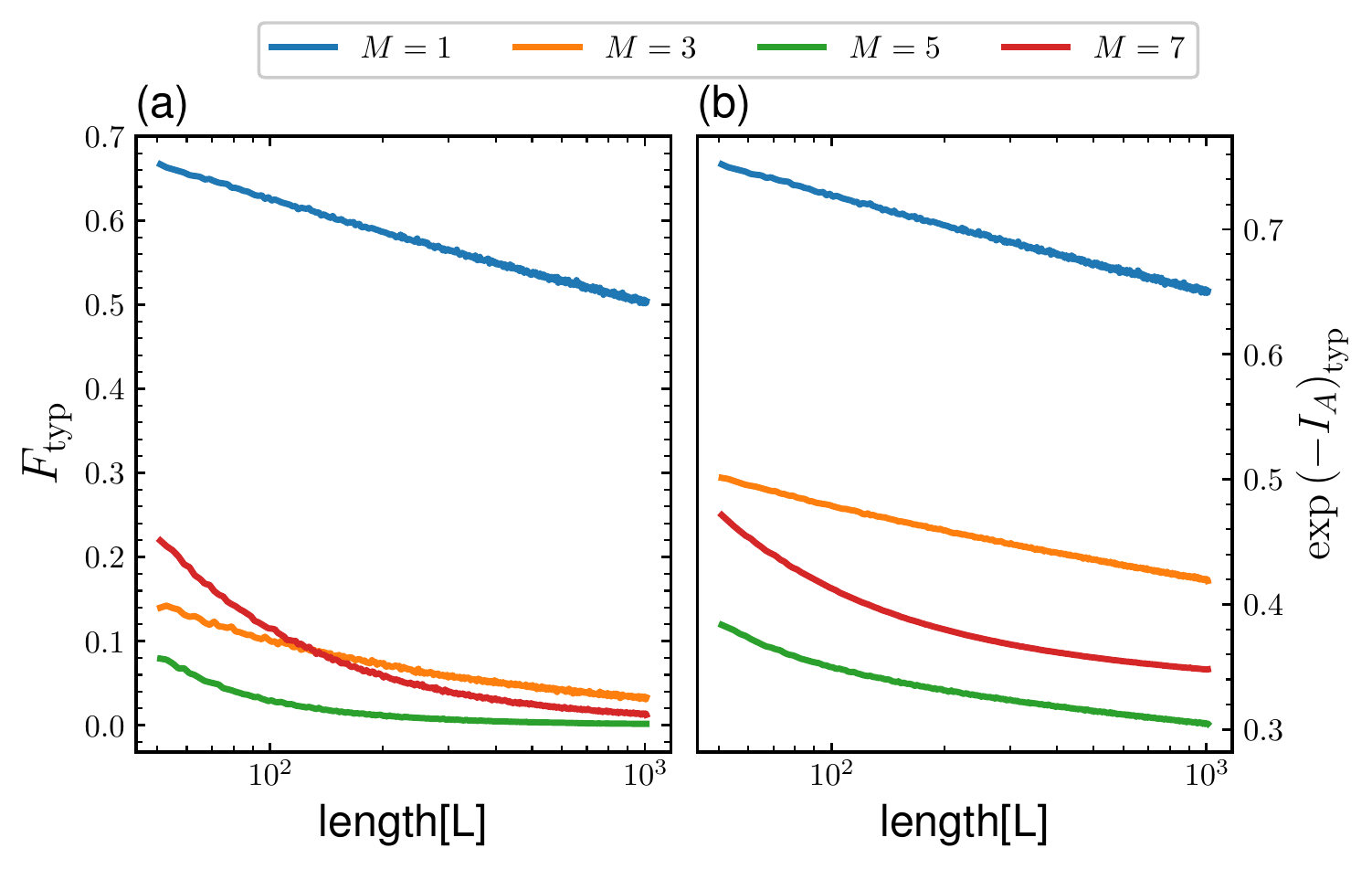}
 \caption{ same as Fig.\ref{tightv01}, but for impurity strength $V_0= 10$ }
\label{tightv10}
\end{figure}

In Figs.\ref{tighBindingFidelity} and \ref{tightv10} we explore the distributed impurity for both weak and strong cases. For the weak strength $V_0=0.1$, we observe fidelity decays  in power law with systems size and becomes more slow with increasing the number of sites which impurity is distributed over on. This in agreement with the fidelity behaviour in the metallic phase of the (extended) AA-model reported in the main text. While for strong impurity, see Fig.\ref{tightv10}, the typical fidelity
decreases with $M$, decaying more fast with a power law of $L$, the larger $M$ is.

 \subsection{Appendix} 
 
In this appendix, we present the average numerical results of fidelity and energy spectrum of the AA model reported in the main text. All comparisons has  given in place in the main text.

\begin{figure*}
\centering
    \includegraphics[width=1.0\textwidth]{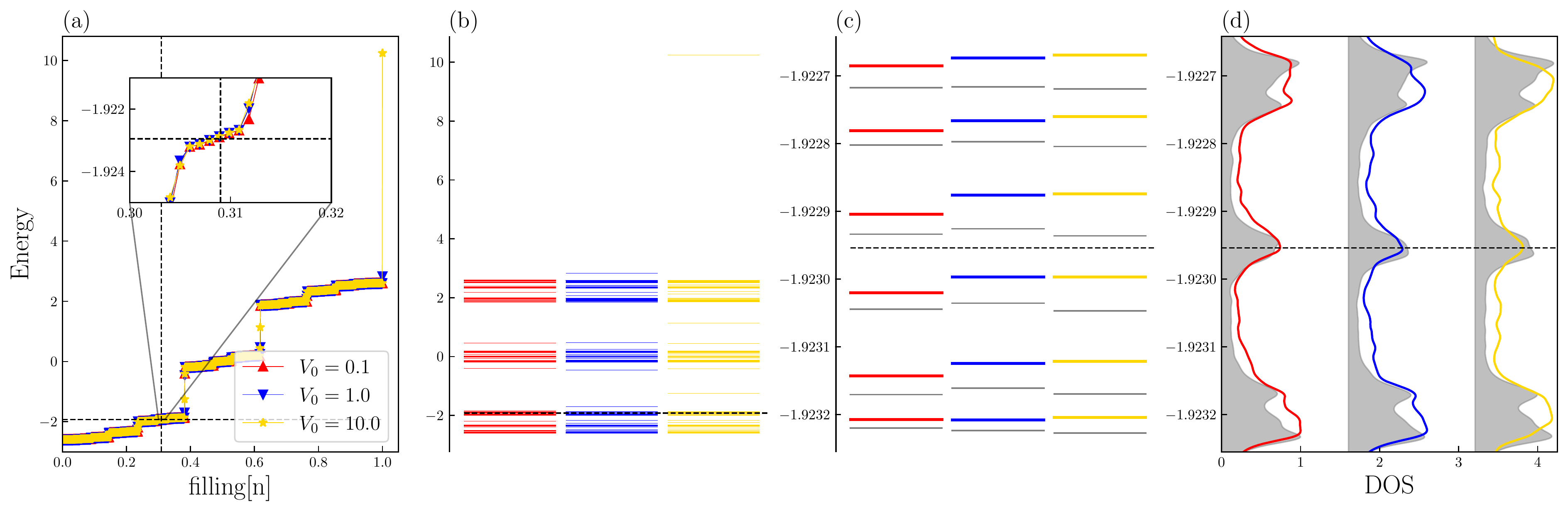}
\caption{Same as Fig~\ref{dos} but with averaging over $200$ realizations. }
\label{dos_ave}
\end{figure*}

\begin{figure*}
\centering
    \includegraphics[width=1.0\textwidth]{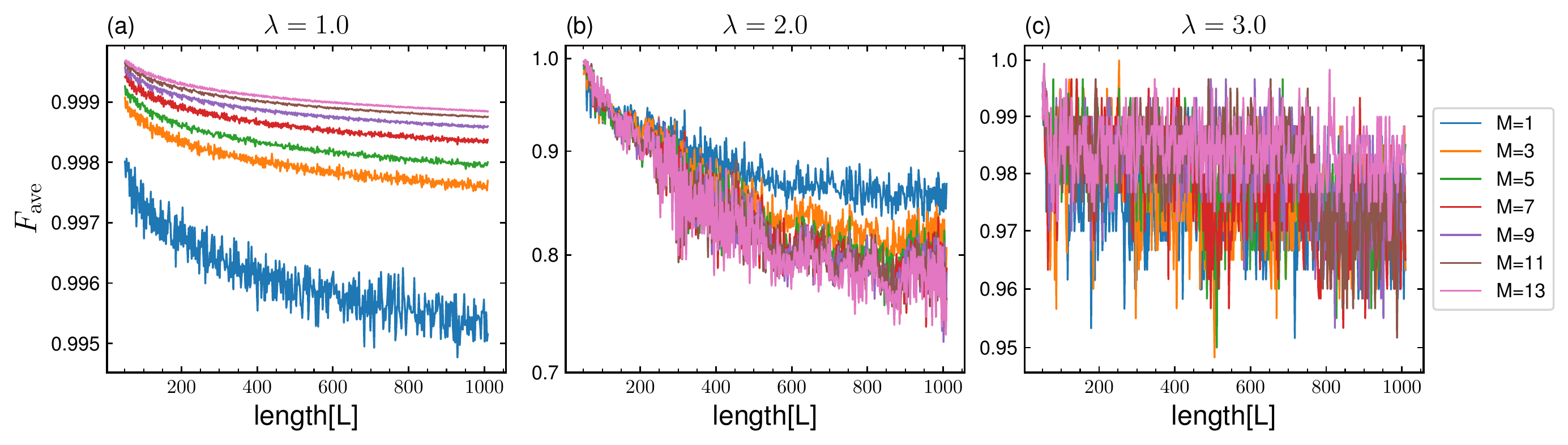}
 \caption{Same as Fig.\ref{single_typ} 
 for a weak impurity
 but for the average value of fidelity $\langle F\rangle$, and data sampled over $1000$ realizations.}
\label{single_ave}
\end{figure*}

\begin{figure*}
\centering
\includegraphics[width=1.0\textwidth]{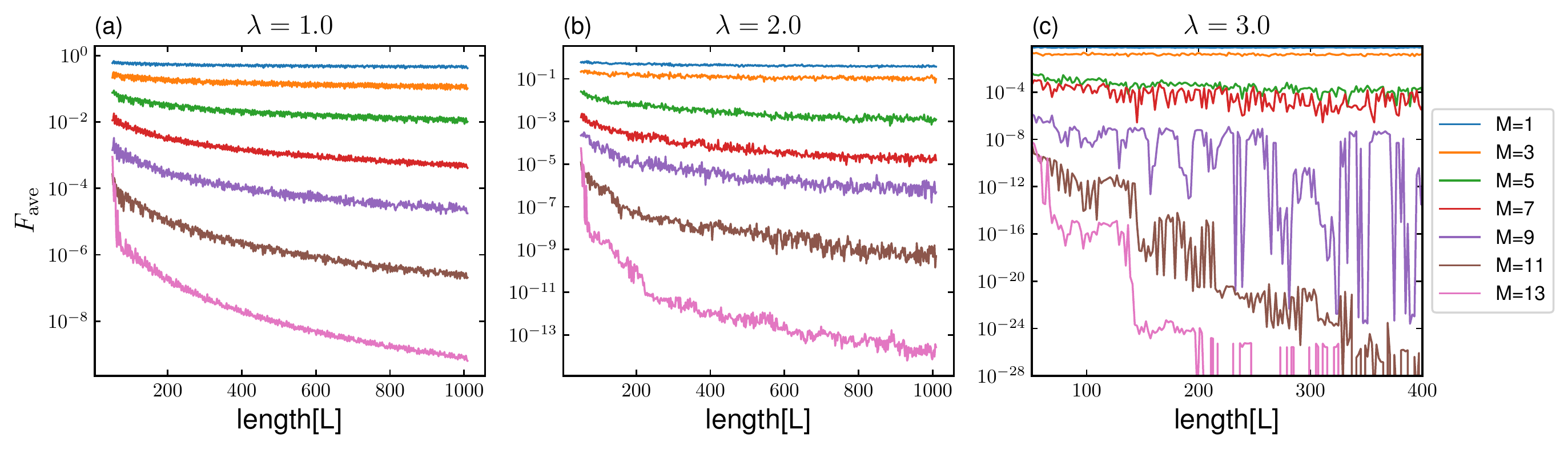}
 \caption{Same as Fig.\ref{multi_typ_v020} but for the average value of fidelity $\langle F\rangle$ with a strong impurity, and data sampled over $1000$ realizations.} 
\label{multi_ave}
\end{figure*}

Fig. \ref{dos_ave} shows the energy diagram and density of states as function of energy $E$, as averaged over the random phases $\phi$ in the Hamiltonian Eq. (\ref{eqEAA}) for $b=0$ and $\lambda=2$ of 200 realizations.  Fig. \ref{single_ave} and Fig\ref{multi_ave} show the average fidelity for multi impurity case for two impurity strength $V_0=0.1$ and $V_0=20$, respectively.

\end{document}